\documentclass[a4paper,11pt, dvipsnames]{article}
\pdfoutput=1 
\DeclareUnicodeCharacter{2032}{\ensuremath{^{\prime}}}
\usepackage{jheppub} 

\usepackage{graphicx}
\usepackage{dcolumn}
\usepackage{bm}
\usepackage{bbm}
\usepackage{amsmath, amsfonts, amsthm, amssymb}
\usepackage{braket}
\usepackage[normalem]{ulem} 
\usepackage{tikz}
\usepackage{rotating}
\usetikzlibrary{decorations.pathmorphing}
\usetikzlibrary{decorations.markings}
\usetikzlibrary{calc}
\usepackage{verbatim}
\usepackage{mathrsfs}
\usepackage{mathtools}
\usepackage[dvipsnames]{xcolor} 
\usepackage{url}
\usepackage{tabularx}
\usepackage{multirow}
\usepackage{subcaption}
\hypersetup{colorlinks=true}
\usepackage[capitalise]{cleveref} 
\usepackage{fontawesome5} 
\usepackage{todonotes}

\usepackage{adjustbox}

\usepackage{mdframed}

\usepackage{xcolor}
\usepackage{tcolorbox}
\usepackage{listings}
\usepackage{newfloat}
\usepackage{algorithm}
\usepackage[noend]{algpseudocode}

\newcommand{\PreComputation}[1]{\State \textbf{Pre-Computation:} #1}
\newcommand{\Construct}[1]{\State \textbf{Construct Basis:} #1}

\newcommand{\SetState}[1]{\State \textbf{Set:} #1}

\newcommand{\Revert}[1]{\State \textbf{Revert:} #1}
\newcommand{\ConstructMatrix}[1]{\State \textbf{Construct Matrix:} #1}
\newcommand{\mycomment}[1]{\textcolor{blue}{$\triangleright$ #1}}

\tcbuselibrary{listings, skins}

\definecolor{codebackground}{RGB}{245,245,245}
\definecolor{codeborder}{RGB}{200,200,200}

\definecolor{blue(munsell)}{rgb}{0.0, 0.5, 0.69}

\newcommand{\dd}{\mathop{}\!\mathrm{d}} 

\begin{document}
\preprint{KA-TP-19-2026}

\title{
Efficient Hamiltonian Truncation: Fast Matrix Construction and Quantum Krylov Diagonalization
}

\author[a]{Rachel Houtz,}
\author[b]{Marco Knipfer,} 
\author[b]{Konstantin Matchev,} 
\author[b]{Alexander Roman,} 
\author[a,c]{Mia West} 

\affiliation[a]{Department of Physics, University of Florida, Gainesville, FL 32611, USA}
\affiliation[b]{Department of Physics and Astronomy, University of Alabama, Tuscaloosa, AL 35487, USA}
\affiliation[c]{Institute for Theoretical Physics, Karlsruhe Institute of Technology, 76131 Karlsruhe, Germany}

\emailAdd{rachel.houtz@ufl.edu}
\emailAdd{mknipfer@ua.edu}
\emailAdd{kmatchev@ua.edu}
\emailAdd{adroman@ua.edu}
\emailAdd{mia.west@kit.edu}

\abstract{
Hamiltonian truncation offers a nonperturbative route to quantum field theory, yet its accuracy is limited by the rapid expansion of the truncated Hilbert space, which drives up computational cost. We tackle this bottleneck with a hybrid strategy that pairs classical and quantum algorithms: 1) we develop an efficient basis-generation scheme built on integer partitions; 2) we speed up the construction of the sparse Hamiltonian matrix using symmetry-aware algorithms; and 3) we explore quantum Krylov diagonalization as a route to the low-lying spectrum. Benchmarking against the free massive scalar and $\phi^4$ theories in two spacetime dimensions, we achieve substantial gains in the computational efficiency of Hamiltonian truncation and chart a path toward future quantum implementations.
}

\maketitle


\section{Introduction}

A major challenge of quantum field theory is obtaining reliable predictions in strongly interacting regimes where standard perturbative techniques fail. Nonperturbative numerical methods provide a path forward, with lattice field theory being the most well-established and successful numerical method.

Hamiltonian truncation is a complementary nonperturbative numerical framework~\cite{Brooks:1983sb, Yurov:1989yu, Yurov:1991my}. Instead of discretizing space, as is done in lattice formulations, Hamiltonian truncation introduces an energy cutoff, thus preserving rotational and translational symmetries and enabling direct access to real-time dynamics. Once a Hamiltonian has been constructed, quantities of interest can be extracted via direct diagonalization. Its successful application to quantum field theories has led to renewed interest in the method within particle physics~\cite{Hogervorst:2014rta, Rychkov:2014eea, Katz:2014uoa}, and results in low dimensions have been demonstrated to be accurate~\cite{Bajnok:2015bgw,  Rychkov:2015vap, James:2017cpc, Anand:2017yij, Anand:2020qnp,EliasMiro:2021aof, Fitzpatrick:2022dwq,Henning:2022xlj,Schmoll:2023eez,Chen:2023glf, Houtz:2025lbv, Li:2026dyb}.

Hamiltonian truncation faces the obstacle that the number of states in the Hilbert space and therefore the size of the Hamiltonian matrix grows exponentially with the energy cutoff. While improvement programs can partially incorporate effects of states above the energy cutoff and successfully reduce errors~\cite{Elias-Miro:2015bqk, Elias-Miro:2017tup, Elias-Miro:2020qwz, Cohen:2021erm,EliasMiro:2022pua, Delouche:2023wsl, Delouche:2024yuo, Demiray:2025zqh, Maestri:2026hqb}, they so far yield only power-law improvement in the cutoff dependence of truncation errors, and therefore cannot confront the underlying exponential growth in computational cost.

Extending Hamiltonian truncation to larger energy cutoffs therefore requires new computational strategies. This has motivated growing interest in applying quantum computing to Hamiltonian truncation as part of the broader effort to develop quantum algorithms for Hamiltonian formulations of quantum field theory~\cite{Kogut:1974ag,Jordan:2011ci, Jordan:2012xnu,Jordan:2014tma, Banuls:2019bmf,Araz:2022tbd,Fromm:2023npm,Abel:2024kuv}. Recently, these ideas have begun to be explored within the Hamiltonian truncation framework, with existing studies demonstrating the ability to model real-time evolution and scattering processes~\cite{Liu:2020eoa,Ingoldby:2024fcy,Ingoldby:2025bdb}. The efficient extraction of energy spectra from increasingly large truncated Hamiltonians remains an important computational challenge for Hamiltonian truncation.

Accordingly, several families of quantum algorithms have been proposed for extracting low-lying spectra from quantum systems, including quantum phase estimation~\cite{Kitaev:1995qy, Nielsen_Chuang_2010, Abrams:1998pd, Aspuru-Guzik:2005zht}, variational quantum eigensolvers~\cite{peruzzo2014variational, McClean:2015vup, Cerezo:2020jpv, Tilly:2021jem}, and quantum Krylov methods~\cite{Parrish:2019ruc, doi:10.1021/acs.jctc.9b01125, PRXQuantum.3.020323, Motta:2019yya, Yoshioka2025, Seki:2021oxk, Cortes:2021esg, Shen:2022lmk}. Among these approaches, quantum Krylov methods are particularly attractive because they avoid large-scale optimization~\cite{Yoshioka2025} while requiring more modest circuit depths than phase estimation~\cite{Cortes:2021esg, Parrish:2019ruc}. Krylov methods have proven particularly effective for extracting eigenvalues from large, sparse matrices~\cite{parlett1998symmetric, doi:10.1137/1.9781611970739,Lanczos:1950zz}.

In this work, we investigate and improve the computational pipeline required to perform large-scale Hamiltonian truncation calculations. We first develop an efficient algorithm for basis-state generation based on integer partitions and present algorithmic improvements that reduce the cost of Hamiltonian construction. We then investigate Krylov-based methods to efficiently extract low-lying spectral information without requiring full diagonalization. Together, these developments provide a computational framework for extending Hamiltonian truncation calculations to larger truncation scales on both classical and quantum computing platforms.

This paper is organized as follows. In Section~\ref{sec:HT}, we review the theoretical framework of Hamiltonian truncation. In Section~\ref{sec:mat-con}, we discuss the construction of the Hamiltonian matrix and present efficient algorithms for basis generation and matrix construction to reduce the cost of the classical stage of the computation. 
In Section~\ref{sec:diag}, we study matrix diagonalization and compare direct diagonalization with the quantum Krylov method using a free 2D scalar theory as a benchmark.
Finally, we apply these methods to the 2D $\phi^2$ and $\phi^4$ theories in Sections~\ref{sec:phi2_results} and~\ref{sec:phi4_results}, and present our numerical results.

\section{Hamiltonian truncation}
\label{sec:HT}

Hamiltonian truncation is a nonperturbative numerical approach that strategically splits the full Hamiltonian of interest into two parts:
	\begin{align}
	H	&=H_0 +  V\,,
    \label{eq:htrunc}
	\end{align}
where $H_0$ is a directly solvable Hamiltonian and $ V$ encodes additional interactions that deform the system from the solved system. Note that the interactions in $V$ can include strong couplings. The Hamiltonian matrix $H$ is written in the basis of states which are eigenstates of $H_0$, each labeled by its eigenvalue $E_i^0$. To make the basis finite-dimensional, one first discretizes the energy spectrum by working in finite volume. Then, a truncation scale $E_{\rm max}$ is introduced and all states with $E_i^0>E_{\rm max}$ are discarded. 

The finite-dimensional matrix $H_\text{eff}$ is  then defined by:
	\begin{align}
	H_{\text{eff}, ij}
		&= E_i^0 \delta_{i,j} +  \bra{E_i^0} V \ket{E_j^0}\,,
	\end{align}
where $\ket{E_i^0}$ are the eigenvectors of $H_0$ with corresponding eigenvalues $E_i^0$ and $\delta_{i,j}$ is the Kronecker symbol. When $V$ includes only relevant deformation operators, the expectation is that the effect of the discarded states with energies above $E_\text{max}$ on the low-lying energy spectra is small. This expectation has borne out under systematic expansion~\cite{Cohen:2021erm, EliasMiro:2022pua, Maestri:2026hqb, Demiray:2025zqh}. Once $H_\text{eff}$ is constructed, the matrix can be directly diagonalized to obtain its spectrum.

\subsection{Free Massive Scalar in 2D}

As a simple toy model, we first explore the free massive scalar in 2D, 
\begin{align}
\mathcal{L} = \frac12 \partial_\mu \phi \partial^\mu \phi - \frac12 m^2 \phi^2 \,. 
    \label{eq:free-scalar}
\end{align}
This theory is exactly solvable, and the eigenstates can be written down directly. We first work in finite volume by quantizing the theory on a spatial circle and imposing boundary conditions
    \begin{align}
    \phi(t, x) = \phi(t, x+ 2 \pi R) \,,
    \end{align}
where $R$ is the radius of the circle. We can then expand $\phi$ in terms of its modes
    \begin{align}
    \phi(t, x)
        &= \frac1{ \sqrt{2 \pi R }}
            \sum_\ell \frac1{ \sqrt {2\omega_\ell}} \left[ a^\dagger_\ell e^{i (\omega_\ell t - \ell x /R)} 
            + a_\ell e^{-i ( \omega_\ell t - \ell x/R)}\right],
    \end{align}
where $\ell$ is an integer and $\omega_\ell \equiv \sqrt{ m_Q^2 + \ell^2/R^2}$. We use the notation $m_Q$ to denote the ``quantization mass," meaning the mass of the free scalar field that defines $H_0$ and the resulting eigenbasis, following the notation of~\cite{Cohen:2021erm}. The raising and lowering operators obey the commutation relation
    \begin{align}
    [a_{\ell}^{{\color{white}\dagger}}\,, a^\dagger_{\ell'}] &= \delta_{\ell, \ell'} \,.
    \end{align}
In terms of raising and lowering operators, 
    \begin{align}
    H_0 = \sum_\ell \omega_\ell a_\ell^\dagger a_\ell \,. 
    \label{eq:H0}
    \end{align}
Imposing a truncation scale $E_{\rm max}$ implicitly forces a maximum $\ell$ value, $\ell_{\rm max}$, such that any state containing particles with larger $\ell$ will necessarily have an energy larger than $E_{\rm max}$. 

Note that because we started with a solvable free massive scalar theory, a simple choice is to identify $m_Q$ with the $m$ in Eq.~(\ref{eq:free-scalar}), yielding a diagonal matrix $H$. One is free, however, to instead arbitrarily separate the mass into two terms:
\begin{equation}
m^2 = m_Q^2 + m_V^2
\label{eq:m2split}
\end{equation}
and write the Hamiltonian as
    \begin{align}
   H  &=  \underbrace{ \displaystyle\int \dd x \ \left[ \frac12 \pi^2 + \frac12 \left( \partial_x \phi\right)^2 + \frac12 m_Q^2 \phi^2 \right] }_{H_0} +  \underbrace{ \displaystyle\int \dd x \  \frac12 m_V^2 \phi^2 }_V
    \end{align}
where $\pi = \partial{\mathcal L}/\partial \dot \phi = \partial_t \phi$ is the momentum conjugate of $\phi$. Following~\cite{Cohen:2021erm}, we call $m_V$ the ``perturbation mass" as it is added as an interaction in $V$. This notation also distinguishes $m_V$ from the quantization mass $m_Q$ which is used to define the basis of states of the quantized theory. Then $H$ can be written in the form of Eq.~(\ref{eq:htrunc}), $H = H_0 + V$, where  
    \begin{align}
    V   &= \displaystyle\int \dd x \ \frac12 m_V^2 :\phi^2: \,,
    \end{align}
where now the $\phi^2$ operator has been normal ordered and an infinite $c$-number has been dropped. 

Splitting the mass $m$ into the quantization and perturbation mass pieces agrees with directly solving $H$ using $m_Q = m$, $m_V=0$ in the $E_{\rm max} \to \infty$ limit. For fixed $R$, schematically, 
    \begin{align}
    H_\text{eff}(m) = H_0( m_Q) + V (m_V)
        \xrightarrow{E_{\rm max} \to \infty} H_0(m) \,,
    \end{align}
with $m$ given by eq.~(\ref{eq:m2split}). We can then compare numerical results of diagonalizing $H_{\rm eff}$ to exact results of $H_0(m)$ and obtain direct access to the truncation errors, as demonstrated in Section~\ref{sec:phi2_results}. This simple case is also useful to benchmark the speed of the $:\phi^2:$ operator construction and diagonalization and check against exact results. Operators of the form $:\phi^2:$ appear, for example, as counterterms resulting from Hamiltonian truncation improvement programs~\cite{Elias-Miro:2015bqk, Elias-Miro:2017tup, Elias-Miro:2020qwz, Cohen:2021erm,EliasMiro:2022pua, Delouche:2023wsl, Delouche:2024yuo, Demiray:2025zqh, Maestri:2026hqb},  and so understanding and decreasing their computational cost is of interest.

\subsection{Scalar \texorpdfstring{$\phi^4$}{phi4} Theory in 2D}

In Section~\ref{sec:phi4_results}, we also consider the interacting 2D $\phi^4$ theory:
\begin{align}
\mathcal{L} = \frac12 \partial_\mu \phi \partial^\mu \phi - \frac12 m^2 \phi^2 - \frac \lambda{4!} \phi^4 \,.
\end{align}
In this case, $H_0$ is defined as above, and our interactions are given by:
    \begin{align}
    V   &= \displaystyle\int \dd x\ \left[ \frac12 m_V^2 :\phi^2: + \frac \lambda{4!} :\phi^4: \right] \,,
    \end{align}
where again $m^2 = m_Q^2 + m_V^2$, see discussion above. 
As the interaction term now has two qualitatively different operators, it is convenient to define
    \begin{align}
    H_{\rm eff} 
        &= H_0 + H_2 + H_4 \,,
    \end{align}
where 
    \begin{align}
    H_2 &\equiv \displaystyle\int \dd x\ \frac12 m_V^2 :\phi^2:\,, 
    \\
     H_4   &\equiv \displaystyle\int \dd x\  \frac \lambda {4!} :\phi^4:\,.
    \end{align}
    
 This theory is superrenormalizable, and so the choice of normal ordering $:\phi^2:$ and $:\phi^4:$ removes all UV divergences. Also, the relevance of the $\phi^4$ interaction, ${\rm dim}(\lambda)=2$, indicates that this theory is well-suited for Hamiltonian truncation, as the relative importance of interactions near $E_{\rm max}$ falls off quickly as $E_{\rm max}$ is increased. 

\section{Matrix Construction}
\label{sec:mat-con}

The first task is the construction of $H_{\rm eff}$. This can be split into two parts. First, one must enumerate the basis of states. This requires finding the eigenstates of $H_0$ with energies $E_i^0$ below the truncation scale $E_{\rm max}$. Second, one must construct the desired matrix operators in the basis of $H_0$.

\begin{figure}[h]
\centering
\includegraphics[width=.45\textwidth, trim= 0cm 0cm 0cm 0.0cm, clip=true]{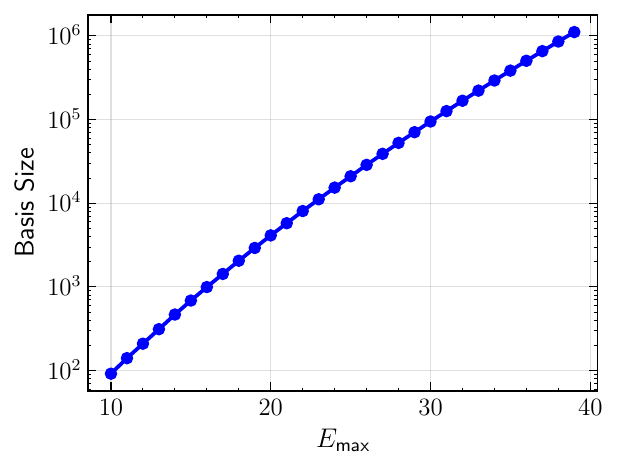}
\caption{Scaling of the basis size with \(E_\text{max}\). This plot is generated with the parameters \(m_Q=1\) and \(R=1\), and includes both odd- and even-occupation number states.}
 
\label{fig:basis_vs_emax}
\end{figure}

Both tasks are challenging because, as demonstrated in~\cref{fig:basis_vs_emax} explicitly for a sample parameter choice, the number of states in the basis grows approximately exponentially with increasing \(E_\text{max}\). It is the severity of this unavoidable scaling that leads to the computational time to construct \(H_\text{\rm eff}\) quickly ballooning for large \(E_\text{max}\).

Fortunately, by constructing algorithms which utilize properties of the effective Hamiltonian, it is possible to significantly temper the severity of the exponential scaling. The algorithms described below are, therefore, constructed to reduce the computational cost of the two steps in comparison to previous methods. All computations in this section were carried out on an AMD Ryzen 9 7950X 16-Core, 64.0 GB DDR5 RAM. The reported execution times correspond to this hardware configuration.

\subsection{Basis Construction}

The operators are represented in the basis of the eigenstates of the free massive scalar theory $H_0$, Eq.~(\ref{eq:H0}). These states correspond to collections of free particles with quantization mass $m_Q$ occupying momentum modes labeled by the integer $\ell$. The normalized Fock states are
\begin{align}
    \ket{ \{n_\ell \} }
    &= \prod_\ell \frac{ ( a_\ell^\dagger )^{n_\ell}} { \sqrt{ n_\ell !}} \ket 0 \,,
\end{align}
where $\ket 0$ is the free vacuum, and $n_\ell$ denotes the occupation number of momentum mode number $\ell$. In this work, we consider the scenario in which the basis is constructed with states that have zero total momentum.

It is convenient to represent these states by their occupancy vectors, 
\begin{align}
\ket{ \{n_\ell \}}
    &= \left|
        \begin{array}{cccccccc}
            \ell &: &0 &-1 &+1 &-2 &+2 &... \\
            n_\ell &: &n_0 &n_{-1} &n_1 &n_{-2} &n_{2} &...
            \end{array}
        \right\rangle \,,
\end{align}
which explicitly encode the particle content of each momentum mode. In 2D, the sign of $\ell$ determines the direction of the momentum.
With this normalization, the ladder operators act on the basis as
    \begin{align}
        a_\ell^\dagger\ket{ ..., n_\ell, ... }
            &= \sqrt{ n_\ell + 1} \ket{ ..., n_\ell + 1, ... }\,,    \\
        a_\ell\ket{ ..., n_\ell, ... }
            &= \sqrt{ n_\ell } \ket{ ..., n_\ell - 1, ... }  \,.
    \end{align}
The occupancy-vector representation makes selection rules and conservation laws explicit, enabling efficient basis generation and filtering.

\begin{figure}[t]
\centering
\includegraphics[width=.75\textwidth, trim=0cm 0cm 0cm 0cm, clip=true]{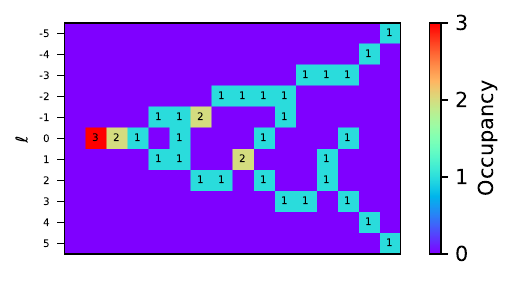} 
\caption{Occupancy-vector representation of truncated basis states for the case of $m_Q=2$, $R=2$, and $E_{\rm max}=7$. Each row represents a momentum mode $\ell$, while each column corresponds to a basis state labeled by the occupancy vector $\{n_\ell\}$. The values of $n_\ell$ are color-coded and printed in the respective boxes. The structure reflects energy constraints and conserved quantum numbers (zero net momentum).
}
\label{fig:basis-occupancy-small}
\end{figure}

\begin{figure}[t]
\centering
\includegraphics[width=.85\textwidth, trim=0cm 0cm 0cm 0cm, clip=true]{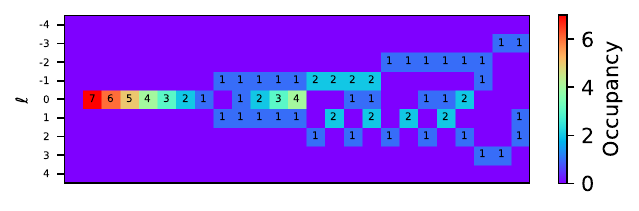}
\includegraphics[width=.85\textwidth, trim=0cm 0cm 0cm 0cm, clip=true]{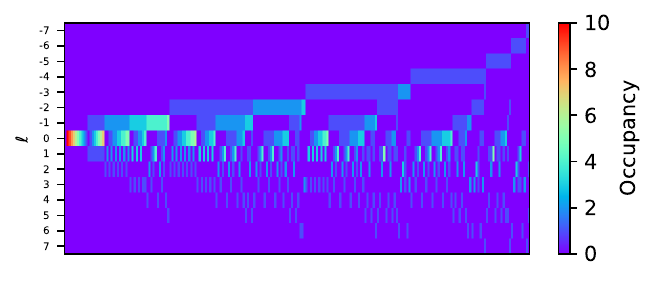} 
\caption{The same as Fig.~\ref{fig:basis-occupancy-small}, but for $m_Q=1$, $R=1$, $E_\text{max}=7$ (top panel) and $m_Q=1$, $R=1.5$, $E_\text{max}=10$ (bottom panel).
This illustrates how the truncated Hilbert space expands and how sparsity and structure evolve with truncation. This representation motivates constraint-aware enumeration strategies and structured matrix construction.
}
\label{fig:basis-occupancy}
\end{figure}

In the occupancy-vector representation, the free-theory energies $E_i^0$ are given by
\begin{equation}
E_i^0
    = E(\{n_\ell\}_i)= \sum_{\ell=-\ell_\text{max}}^{\ell_\text{max}} n_\ell\,
\sqrt{m_Q^2 + \frac{\ell^2}{R^2}}
    = \sum_{\ell=-\ell_\text{max}}^{\ell_\text{max}} n_\ell \omega_\ell\,.
\end{equation}
Fig.~\ref{fig:basis-occupancy-small} visualizes this representation as a heat map of occupancies $n_\ell$ per momentum mode $\ell$ across the truncated basis for a low $E_{\rm max}$ benchmark ($\ell_\text{max}=5$), illustrating the highly constrained subset of Fock space selected by energy and symmetry constraints. Fig.~\ref{fig:basis-occupancy} shows the same visualization as the mass $m_Q$ is lowered (top panel) or the energy cutoff $E_\text{max}$ increased (bottom panel), emphasizing how the occupied region and sparsity patterns evolve with $m_Q$, $R$ and $E_{\rm max}$. 

One can also recast basis enumeration as a constraint-satisfaction problem on integer-valued occupancy vectors, enabling efficient classical filtering and providing a natural interface for quantum and quantum-inspired algorithms that exploit structured state spaces.
Requiring that each state in the basis has zero total momentum allows us to construct the basis rapidly by considering successively increasing integer partitions. The integer partitions are defined as follows: given an integer $\ell$, the partitions are the set of all possible combinations of smaller integers that sum to $\ell$. Each integer partition is considered to be a set of momentum modes of a particular sign, which can be combined with another of opposite sign to construct a state in the basis with zero net momentum. This procedure is described in more detail in the pseudocode in Algorithm~\ref{alg:basis_construction}. Note that at this stage it is beneficial to record which pairs of states in the basis are exactly anti-symmetric under \(\mathbb{Z}_2\)-parity. This is because the entries in the transition matrix for these pairs are identical, thus there is the possibility to increase computational efficiency by computing the transition amplitudes for only one of the pair. This property was also recognized and utilized by \cite{Rychkov:2014eea}.

\begin{algorithm}
\fontsize{10pt}{14pt}\selectfont
\caption{\textbf{Basis Construction Algorithm.} \newline The required inputs are \(m_Q,\,R\) and \( E_\text{max}\). The basis is returned, optionally along with a list of \(\mathbb{Z}_2\)-parity antisymmetric pairs.}\label{alg:basis_construction}
\begin{algorithmic}
\PreComputation
\State \mycomment{Compute the effective maximum momentum.}
\State \(\ell_\mathrm{max, eff} =  \text{floor}\left[R\sqrt{E_\text{max}^2/4-m_Q^2} \right]\)
\Statex \vspace{-1em}
\Construct
\For{\(\ell=0,...,\ell_\mathrm{max, eff}\)}:
\State\mycomment{Generate integer partitions up to the effective maximum momentum.}
    \State \texttt{partitions}[\({\ell}\)] = all integer partitions of \({\ell}\)
    \Statex \vspace{-1.25em}
    \For{partition \(P\) in \texttt{partitions}[\({\ell}\)]}:

        \For{partition \(Q\) in \texttt{partitions}[\({\ell}\)]}:
        \State \mycomment{Take the union of the two partitions \(P\) and \(Q\).}
            \State \texttt{candidate\_state} = \(P\cup (-Q)\) 
            \Statex \vspace{-1.25em}
            \State \mycomment{Confirm the state has less energy than \(E_\text{max}\).}
            \If{\texttt{energy}(\texttt{candidate\_state})\(<E_\text{max}\)}
            \State \(n_0=0\)

            \State \mycomment{Add additional \(\ell=0\) modes until \(E_\text{max}\) is reached.}
            \While{\texttt{energy}\,(\texttt{candidate\_state})\(<E_\text{max}\)}
            \State Add \texttt{candidate\_state} to \texttt{basis}
            \State\texttt{candidate\_state}\(\:[n_0\rightarrow n_0+1]\)
            \State \(n_0+=1\)
            \EndWhile
            \Statex \vspace{-1.25em}
            \State \mycomment{(Optional) Record \(\mathbb{Z}_2\)-antisymmetric pairs.}
            \If{\(P\neq Q\)}
                \State Add \texttt{candidate\_state\_index} to \texttt{Z2\_antisymmetric\_pairs} 
            \EndIf
        \EndIf
\EndFor
\EndFor
\EndFor
\Statex \vspace{-1.25em}
\State \Return \texttt{basis}
\State \mycomment{(Optional) Or additionally return \(\mathbb{Z}_2\)-antisymmetric pairs.}
\State \Return \texttt{basis, Z2\_antisymmetric\_pairs}
\end{algorithmic}
\end{algorithm}

\begin{figure}[h]
\centering
\includegraphics[width=.66\textwidth, trim= 0cm 0cm 0cm 0.0cm, clip=true]{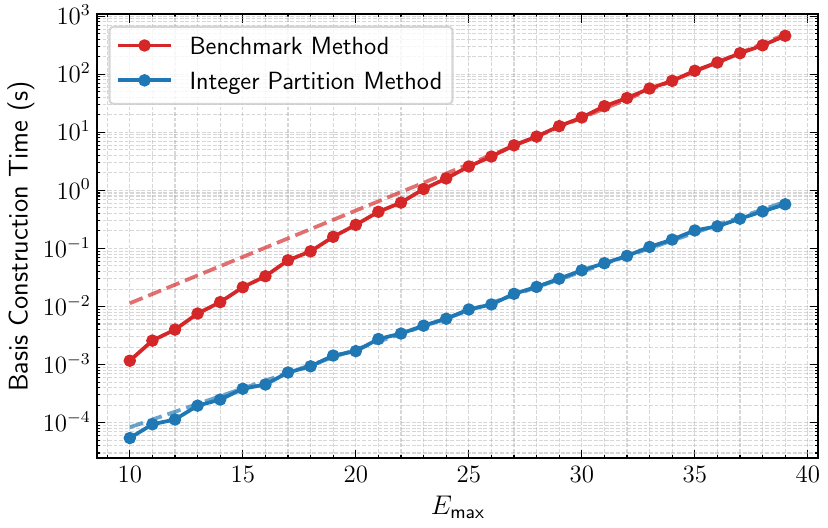}
 \caption{Timing comparison between the basis generation code used in \cite{Cohen:2021erm} (red) as the ``Benchmark Method" and the ``Integer Partition Method" described in this work (blue). This plot is generated with parameters $m_Q=1,R=1$, for varied \(E_\text{max}\). The scaling of the number of states in the basis with \(E_\text{max}\) can be seen in \cref{fig:sparsity_scaling}. Note that for comparison, we re-generate all integer partitions for each data point. A linear fit for \(E_\text{max}\in[25,40)\) to the ``Benchmark Method" has the form \(\log(\tfrac{t}{1s})=0.16\,(E_\text{max})-3.54\), and to the ``Integer Partition Method" from \(E_\text{max}\in[15,40)\) the form \(\log(\tfrac{t}{1s})=0.13\,(E_\text{max})-5.42\), where \(t\) is the basis construction time. The differing starting values for the linear fit are chosen as the respective scaling regimes are approximately reached.}
\label{fig:basis-size}
\end{figure}

An implementation of this algorithm in \texttt{Python} can be found in the following  \href{https://github.com/miarobin/FasterHT}{GitHub repository \faGithub}. Fig. \ref{fig:basis-size} shows the timing comparison to what we call the ``benchmark" method \cite{Cohen:2021erm}. The benchmark method generates the basis recursively by successively adding modes of decreasing momentum until the energy cut-off \(E_\text{max}\) is reached. States which do not have zero total momentum are then discarded. 
Compared to the benchmark method (shown in red), the Integer Partition Method (shown in blue) performs significantly better and with an improved exponential scaling.

\subsection{Identify nonzero entries}

\begin{figure}[htbp]
    \centering

    \begin{subfigure}{\textwidth}
        \centering
        \includegraphics[width=.45\textwidth, trim= 0cm 0cm 0cm 0.0cm, clip=true]{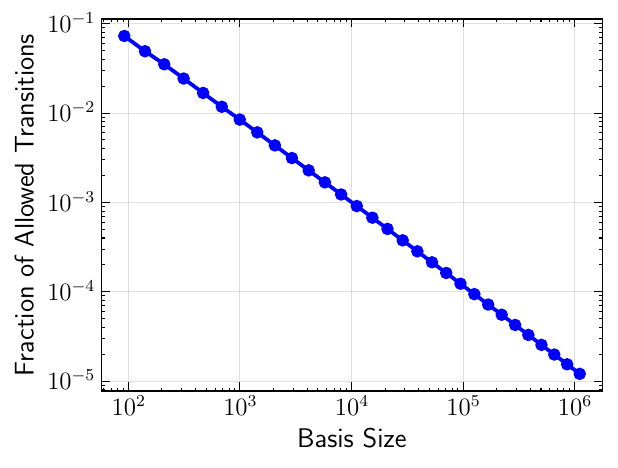}
        \includegraphics[width=.45\textwidth, trim= 0cm 0cm 0cm 0.0cm, clip=true]{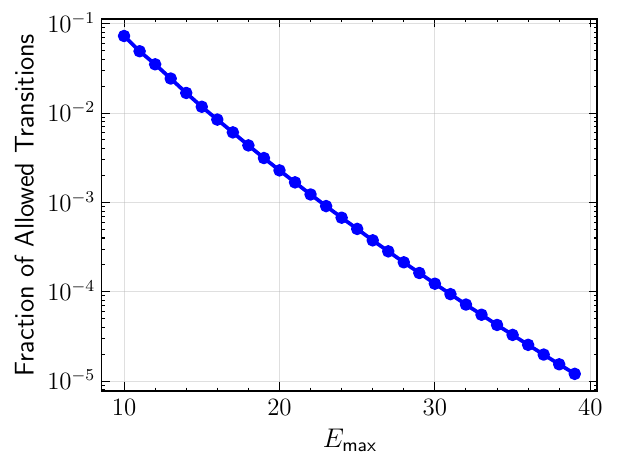}
        \caption{2D \(\phi^2\) Sparsity Scaling.}
        \label{fig:phi2sparsity}
    \end{subfigure}
    \hfill
    \begin{subfigure}{\textwidth}
        \centering
        \includegraphics[width=.45\textwidth, trim= 0cm 0cm 0cm 0.0cm, clip=true]{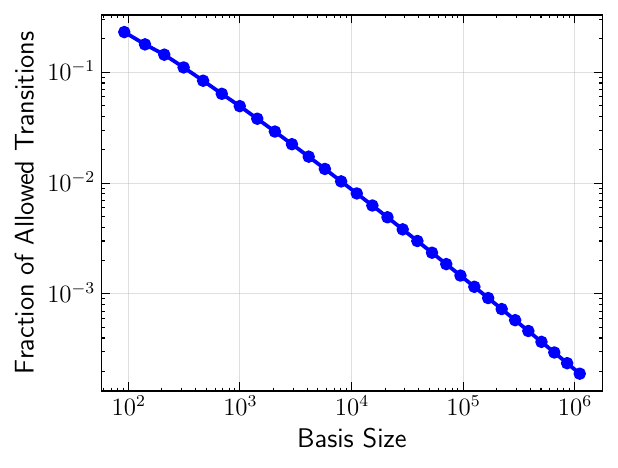}
        \includegraphics[width=.45\textwidth, trim= 0cm 0cm 0cm 0.0cm, clip=true]{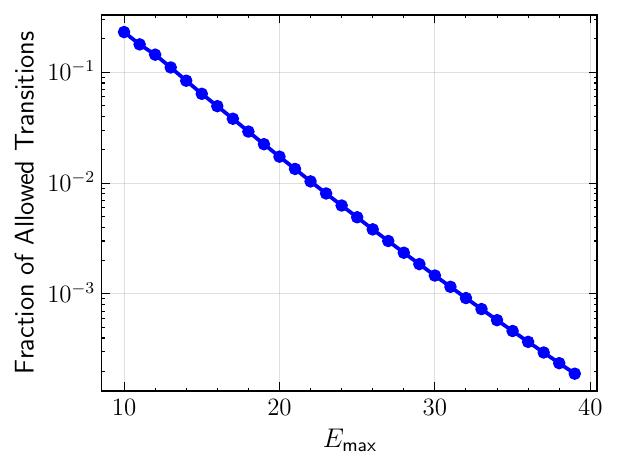}
        \caption{2D \(\phi^4\) Sparsity Scaling.}
        \label{fig:phi4sparsity}
    \end{subfigure}

    \caption{The sparsity scaling, specifically fraction of allowed transitions, with the basis size (left) and \(E_\text{max}\) (right). This plot is generated with the parameters \(m_Q=1.0\) and \(R=1.0\), and includes both odd- and even-occupation number states.}
    \label{fig:sparsity_scaling}
    
\end{figure}

\begin{figure}[t]
\centering
\includegraphics[width=.48\textwidth, trim=0cm 0cm 0cm 0cm, clip=true]{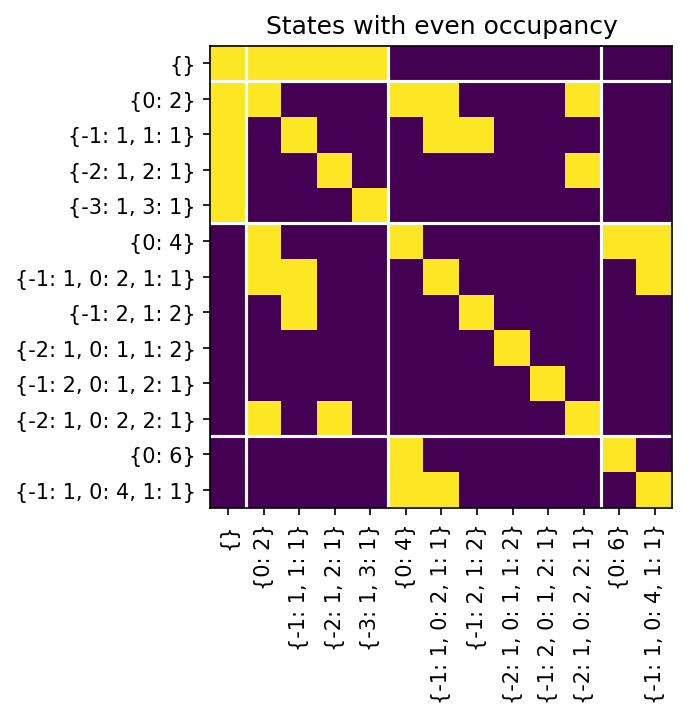} 
\includegraphics[width=.48\textwidth, trim=0cm 0cm 0cm 0cm, clip=true]{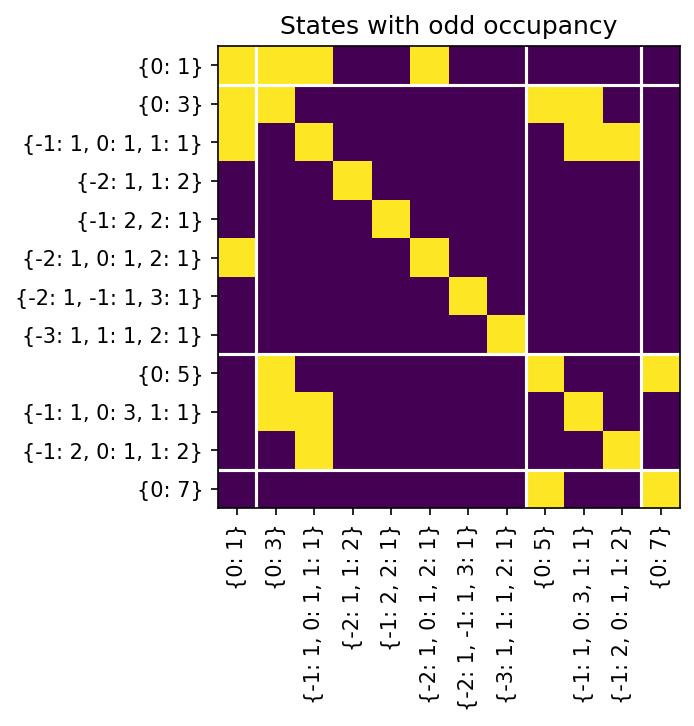}
\caption{The filling pattern of the transition matrix $\bra{E_i} V \ket{E_j}$ for the case of $m_Q = m_V =1$, $R=1$, and $E_\text{max}=7$.  The 25 allowed states depicted in the top panel in Fig.~\ref{fig:basis-occupancy} can be grouped into 13 even-occupancy states (left panel) and 12 odd-occupancy states (right panel). Yellow (purple) entries indicate allowed (forbidden) transitions.
}
\label{fig:adjacency}
\end{figure}

Quick identification of the nonzero entries for the matrix is critical for efficient matrix construction.
The size of the basis grows exponentially with the truncation scale $E_{\rm max}$, see Fig.~\ref{fig:basis_vs_emax}. Fortunately, raising $E_{\rm max}$ also results in ever more sparse interaction Hamiltonians, as demonstrated in Fig.~\ref{fig:sparsity_scaling}. Only a vanishing fraction of matrix elements must be calculated. A visualization example of the position of the nonzero matrix entries is shown in Fig.~\ref{fig:adjacency}, which groups the 25 allowed states depicted in the top panel in Fig.~\ref{fig:basis-occupancy} into 13 even-occupancy states (left panel) and 12 odd-occupancy states (right panel). Yellow (purple) entries indicate allowed (forbidden) transitions. Transitions between an even-occupancy state (left panel) and an odd-occupancy state (right panel) are not allowed, which greatly simplifies the construction. For clarity, the rows and columns in Fig.~\ref{fig:adjacency} are explicitly labeled by the respective occupancy vectors $\{n_\ell\}$ of the basis states, using the python dictionary syntax \{{\rm ``key"}: {\rm ``value"}\}.

The number of possible mode creations and annihilations is fixed by the respective operator, and therefore only transitions changing the total number of modes in a state, termed occupation number, by either 0 or 2 for the case of \(H_2\) or additionally \(4\) in the case of \(H_4\), are feasible. Furthermore, only those transitions preserving momentum between the initial and final states are permitted. Imposing both such requirements for the \(H_2\) operator results in the filling pattern shown in \cref{fig:adjacency} for a sample parameter set.

\subsection{Filling in nonzero entries}

The matrix element for an initial state $\ket{a}$ which transitions to a final state $\ket{b}$ is detailed in \cref{tab:H2matrixelements} for the case of the $H_2$ operator and in \cref{tab:H4matrixelements} for the $H_4$ operator. Without loss of generality we take \(E_a\geq E_b\). Additionally, both the initial and final states must have identical total momentum which in this work we always consider to be zero. The falling factorial notation used in \cref{tab:H2matrixelements} and \cref{tab:H4matrixelements} is defined as:
\begin{align}\label{eq:falling-factorial}
    (m)_n = \frac{m!}{(m-n)!}.
\end{align} 
We note that $\Delta_{\ell_i}$ can equal zero, in which case the falling factorial $(n_{\ell_i})_{\Delta_{\ell_i}}=(n_{\ell_i})_0=1$, so long as the other conditions are obeyed.

\begin{table}
\centering
\renewcommand{\arraystretch}{1.7}
\renewcommand{\tabularxcolumn}[1]{m{#1}}
\begin{tabularx}{.9\textwidth} { 
  | >{\centering\arraybackslash}c 
  | >{\raggedright\arraybackslash}X 
  | >{\raggedright\arraybackslash}X | }
\hline
\begin{tabular}{c}
Rescaled Matrix Element: \\
\(\left(\dfrac{1}{m_V^2}\right)\bra{b}H_2\ket{a}\)
\vspace{0.2em}
\end{tabular} &
Occupation Number Difference in $\ket{b}$ & Momentum Conserving Conditions\\
\hline
\(\displaystyle
\frac{\sqrt{(n_{\ell_1})_{\Delta_{\ell_1}} (n_{\ell_2})_{\Delta_{\ell_2}}}}
{\Delta_{\ell_1}!\Delta_{\ell_2}!\sqrt{\Omega_{\ell_1}\Omega_{\ell_2}}}
\)
&
\(n_{\ell_1}\rightarrow n_{\ell_1}-\Delta_{\ell_1}\), \(n_{\ell_2}\rightarrow n_{\ell_2}-\Delta_{\ell_2}\).
&
\(\sum_{i=1}^2 \Delta_{\ell_i}\ell_i = 0\), and \(\sum_{i=1}^2 \Delta_{\ell_i}=2\).
\\

\hline

$\displaystyle
\sum_{l=-\ell_\text{max}}^{+\ell_\text{max}}\frac{n_\ell}{\omega_\ell},
$
&
$\ket{b}=\ket{a}$.
&

\\

\hline
\end{tabularx}
\caption{Non-zero matrix elements of the free Hamiltonian $H_2$, for a transition between initial state \(\ket{a}\) and final state \(\ket{b}\). Without loss of generality, the energy of state \(\ket{a}\) is greater than that of \(\ket{b}\). An occupation number cannot fall below zero, as this would correspond to annihilation of the vacuum which has zero amplitude. Note the occupation numbers in the first column are for state \(\ket{a}\). We additionally require that \(\ell_1\neq\ell_2\), and \(\Delta_{\ell_i}\in\mathbb{Z},\,\Delta_{\ell_i}\geq0\), where $\Delta_{\ell_i}$ indicates the decrease in occupancy of mode $\ell_i$, see the central column. Repeated operators are accomplished by setting combinations of \(\Delta_{\ell_i}=0\). This table is organized by the number of annihilation and creation operators in the following way: the first row is the effect of two annihilation operations, the second from one annihilation and one creation operator, mirroring the operation separation as it is in the \href{https://github.com/miarobin/FasterHT}{GitHub repository \faGithub}. Note that \(\Omega_{\ell_i}\) is defined as \(\omega_{\ell_i}^{\Delta_{\ell_i}}\) and the falling factorial $(m)_n$ is defined in \cref{eq:falling-factorial}.
}
\label{tab:H2matrixelements}

\end{table}

\begin{table}
\centering
\renewcommand{\arraystretch}{1.5}
\renewcommand{\tabularxcolumn}[1]{m{#1}}
\begin{adjustbox}{minipage=16.25cm, center}
\begin{tabularx}{\textwidth}{
  | >{\centering\arraybackslash}c
  | >{\hsize=0.8\hsize\raggedright\arraybackslash}X
  | >{\hsize=1.2\hsize\raggedright\arraybackslash}X |
}
\hline
\begin{tabular}{c}
Rescaled Matrix Element: \\
\(\left(\dfrac{8\pi R}{\lambda}\right)\bra{b}H_4\ket{a}\)
\vspace{0.2em}
\end{tabular} 

 &
Occupation Number Difference in $\ket{b}$ & Momentum Conserving Conditions\\
\hline
\(\displaystyle
\frac{\sqrt{(n_{\ell_1})_{\Delta_{\ell_1}}
(n_{\ell_2})_{\Delta_{\ell_2}}
(n_{\ell_3})_{\Delta_{\ell_3}}
(n_{\ell_4})_{\Delta_{\ell_4}}}}
{\Delta_{\ell_1}!\Delta_{l_2}!\Delta_{\ell_3}!\Delta_{\ell_4}!
\sqrt{\Omega_{\ell_1}\Omega_{\ell_2}\Omega_{\ell_3}\Omega_{\ell_4}}}
\)
&
For \(i=1,\dots,4\): \(n_{\ell_i}\rightarrow n_{\ell_i}-\Delta_{\ell_i}\) 
&
\(\sum_{i=1}^4\Delta_{\ell_i}\ell_i=0\), and 
\(\sum_{i=1}^4\Delta_{\ell_i}=4\).
\\

\hline

\(\displaystyle
\frac{\sqrt{(n_{\ell_1})_{\Delta_{\ell_1}}
(n_{\ell_2})_{\Delta_{\ell_2}}
(n_{\ell_3})_{\Delta_{\ell_3}}
(n_{\ell_4}+1)}}
{\Delta_{\ell_1}!\Delta_{\ell_2}!\Delta_{\ell_3}!
\sqrt{\Omega_{\ell_1}\Omega_{\ell_2}\Omega_{\ell_3}\omega_{\ell_4}}}
\)
&
For \(i=1,\dots,3\):
$n_{\ell_i}\rightarrow
n_{\ell_i}-\Delta_{\ell_i}$,
$n_{\ell_4}\rightarrow n_{\ell_4}+1$.
&
$\sum_{i=1}^3\Delta_{\ell_i}\ell_i-\ell_4=0,$ and

$\sum_{i=1}^3\Delta_{\ell_i}=3$.
\\

\hline

\(\displaystyle
\frac{\sqrt{n_{\ell_1}n_{\ell_2}(n_{\ell_3}+1)(n_{\ell_4}+1)}}
{\sqrt{\omega_{\ell_1}\omega_{\ell_2}\omega_{\ell_3}\omega_{\ell_4}}}
\)
&
$n_{\ell_{1,2}}\rightarrow n_{\ell_{1,2}}-1$, 

$n_{\ell_{3,4}}\rightarrow n_{\ell_{3,4}}+1$.
&
For \(i=1,\dots,\text{ceil}\big[\frac{\ell_1-\ell_2}{2}\big]-1\) where w.l.o.g.
\(\ell_1>\ell_2\):

\(\ell_3=\ell_1-i\),\;\(\ell_4=\ell_2+i\) and \(\ell_3\neq\ell_4\).

\\

\hline

\(\displaystyle
\frac{\sqrt{n_{\ell_1}n_{\ell_2}(n_{\ell_3}+1)(n_{\ell_3}+2)}}
{2\sqrt{\omega_{\ell_1}\omega_{\ell_2}\omega_{\ell_3}^2}}
\)
&
$n_{\ell_{1,2}}\rightarrow n_{\ell_{1,2}}-1$, 

$n_{\ell_{3}}\rightarrow n_{\ell_{3}}+2$.
&
If \(\big[\frac{\ell_1-\ell_2}{2}\big]\in\mathbb{Z}\) where w.l.o.g.
\(\ell_1>\ell_2\):

\(\ell_3=(\ell_1+\ell_2)/2\).

\\

\hline

\(\displaystyle
\frac{\sqrt{(n_{\ell_1})_{\Delta_{\ell_1}}
(n_{\ell_2})_{\Delta_{\ell_2}}}}
{\Delta_{\ell_1}!\Delta_{\ell_2}!
\sqrt{\Omega_{\ell_1}\Omega_{\ell_2}}}
\sum_{\ell=-\ell_{\max}}^{\ell_{\max}}
\frac{n_\ell-\sum_{i=1}^2\delta_{\ell,\ell_i}\Delta_{\ell_i}}
{\omega_\ell(\sum_{i=1}^2\delta_{\ell,\ell_i}\Delta_{\ell_i}+1)}
\)
&
For \(i=1,2\):
$n_{\ell_i}\rightarrow n_{\ell_i}-\Delta_{\ell_i}$.
&
$\sum_{i=1}^2\Delta_{\ell_i}\ell_i=0,$ and

$\sum_{i=1}^2\Delta_{\ell_i}=2$.
\\

\hline

\(\displaystyle
\sum_{\ell=-\ell_{\max}}^{+\ell_{\max}}
\left(
\frac{n_\ell(n_\ell-1)}{4\omega_\ell^2}
+\sum\limits_{\ell'=\ell+1}^{+\ell_\text{max}}
\frac{n_\ell n_{\ell'}}
{\omega_\ell\omega_{\ell'}}
\right)
\)
&
$\ket{b}=\ket{a}$.
&

\\

\hline
\end{tabularx}
\end{adjustbox}
\caption{Non-zero matrix elements of the quartic interaction Hamiltonian $H_4$, for a transition between initial state \(\ket{a}\) and final state \(\ket{b}\).
Without loss of generality, the energy of state \(\ket{a}\) is greater than that of \(\ket{b}\).
Note the occupation numbers in the first column are for state \(\ket{a}\).
An occupation number cannot fall below zero, as this would correspond to annihilation of the vacuum which has zero amplitude.
Note that \(\Delta_{\ell_i}\in\mathbb{Z},\,\Delta_{\ell_i}\geq0\), where $\Delta_{\ell_i}$ indicates the decrease in occupancy of mode $\ell_i$, see the central column.
We additionally require that all $\ell_i$ are different from one another, and instead repeated operators are accomplished by setting combinations of \(\Delta_{\ell_i}=0\). This table is organized by the number of annihilation and creation operators in the following way: the first row is the effect of four annihilation, the second and fifth are the effect of three annihilation and one creation, and the remaining rows from two annihilation and two creation operators. This mirrors the operation separation as it is in the \href{https://github.com/miarobin/FasterHT}{GitHub repository \faGithub}. Note \(\Omega_{\ell_i}\) is defined as \(\omega_{\ell_i}^{\Delta_{\ell_i}}\) and the falling factorial $(m)_n$ is defined in \cref{eq:falling-factorial}.
}
\label{tab:H4matrixelements}
\end{table}

\begin{figure}[htb]
\centering
    \begin{subfigure}{\textwidth}
        \centering
        \includegraphics[width=.6\textwidth, trim=0cm 0cm 0cm 0cm, clip=true]{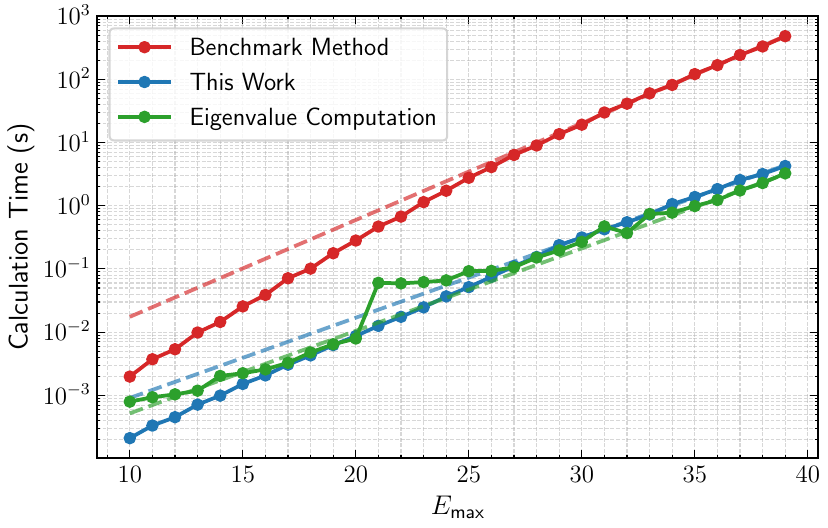} 
        \caption{2D \(\phi^2\) calculation total timing, run with \(m_V=1.0\). A linear fit for \(E_\text{max}\in[30,40)\) to the benchmark method has yields \(\log(\tfrac{t}{1s})=0.15\,(E_\text{max})-3.29\), and to the method from this work from \(E_\text{max}\in[30,40)\) yields \(\log(\tfrac{t}{1s})=0.13\,(E_\text{max})-4.31\), where \(t\) is the transition matrix construction time. A fit to the time to compute the five lowest eigenvalues for \(E_\text{max}\in[35,40)\) yields \(\log(\tfrac{t}{1s})=0.13\,(E_\text{max})-4.59\), where \(t\) is the respective calculation time.}
        \label{fig:phi2summary}
    \end{subfigure}
    \hfill
    \begin{subfigure}{\textwidth}
        \centering
        \includegraphics[width=.6\textwidth, trim=0cm 0cm 0cm 0cm, clip=true]{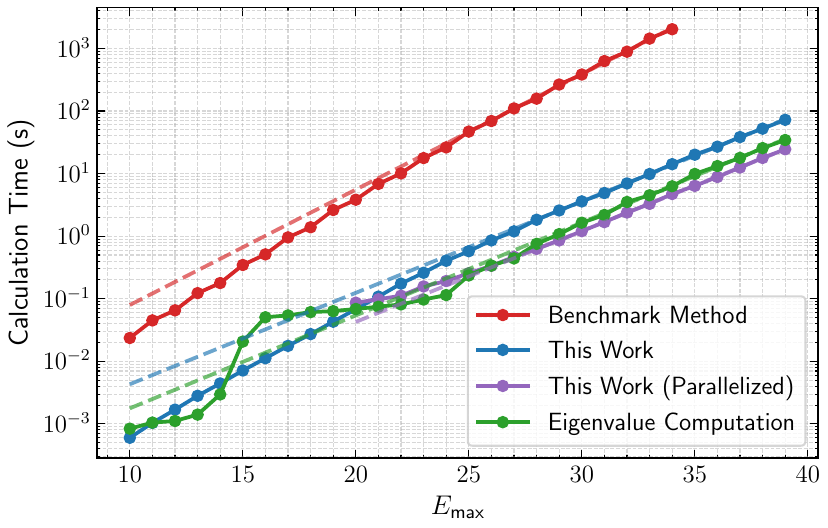} 
        \caption{2D \(\phi^4\) calculation total timing, with \(\lambda=4\pi\). A linear fit for \(E_\text{max}\in[25,30)\) to the benchmark method yields \(\log(\tfrac{t}{1s})=0.18\,(E_\text{max})-2.95\), and to the method described in this work from \(E_\text{max}\in[30,40)\) yields \(\log(\tfrac{t}{1s})=0.15\,(E_\text{max})-3.83\). A linear fit to the time to compute the five lowest eigenvalues for \(E_\text{max}\in[30,40)\) yields \(\log(\tfrac{t}{1s})=0.15\,(E_\text{max})-4.24\). The parallelized version of the total matrix construction methods described in this work run on four CPU cores for \(E_\text{max}\in[30,40)\) yields \(\log(\tfrac{t}{1s})=0.15\,(E_\text{max})-4.27\).}
        \label{fig:phi4summary}
    \end{subfigure}
\caption{
Summary of the total calculation time for transition matrix of the even sector only, for the algorithm described in this work (blue) compared with that from \cite{Cohen:2021erm} termed the benchmark method (red). The total construction time includes basis and matrix construction, in addition to splitting the states into even and odd sectors. For the method described in this work, a hashable basis structure is required which is also included in the construction time. In \cref{fig:phi2summary} is the 2D \(\phi^2\) scenario and in \cref{fig:phi4summary} is the 2D \(\phi^4\) scenario, which also includes a parallelized version of the matrix construction algorithm (purple). Included in these figures is the time for computing the five smallest eigenvalues of the transition matrix (green). The plot is generated with \(m_Q=1.0\) and \(R=1.0\).
     The differing starting values of \(E_\text{max}\) for the linear fits are chosen such that the scaling regimes are reached. The `jumps' in time to calculate the eigenvalues are hardware dependent and therefore likely due to \texttt{scipy}'s internal mechanisms for optimization.}
\label{fig:timingsummary}
\end{figure}

The transition matrices are extremely sparse, which can be utilized to improve computational efficiency. We achieve this by looping only once through all states in the basis, and identifying exactly which transitions are permitted. As there are very few, this can be done quickly. The location of the matrix element can then be found efficiently on a computer by looking-up the final-state in a hashable data structure storing the basis states.\footnote{Note that there is some time-cost associated with creating the hashable data structure to store the basis. In the \href{https://github.com/miarobin/FasterHT}{GitHub repository \faGithub} we implement this simultaneously to splitting the basis into even and odd occupancy states which we also perform for the benchmark method, which is included in \cref{fig:timingsummary}. This is a negligibly low-cost operation in comparison to matrix construction.} The matrix element is then filled considering the form of \(V\): specifically from selecting the appropriate formula from \cref{tab:H2matrixelements} for the \(H_2\) operation and \cref{tab:H4matrixelements} for the \(H_4\) operation.

In addition, if antisymmetric pairs of states under \(\mathbb{Z}_2\)-parity are recorded during basis construction, it is possible to only loop through one of each pair. 
This is because the transition amplitudes are identical, which can be attributed to the operators \(H_0, H_2\), and \(H_4\)  being symmetric under \(\mathbb{Z}_2\)-parity. It is then simple to insert the amplitudes for the other member of the pair into the transition matrix, while taking care to ensure the matrix elements corresponding to transitions e.g.\ between two antisymmetric \(\mathbb{Z}_2\)-parity pairs are inserted correctly. For small values of \(E_\text{max}\), mostly \(\mathbb{Z}_2\)-parity symmetric states exist. However, as the number of integer partitions grows, the number of \(\mathbb{Z}_2\)-parity antisymmetric states increases. In the limit of large \(E_\text{max}\) up to a factor of two, improvement in computation time could be gained in principle. 

As already mentioned, without loss of generality, we select that the initial state is of greater or equal energy to the final state in the transition. The transpose of the matrix corresponds to the opposite case. Finally, even and odd occupancy states are considered separately as transitions between them are forbidden. For the execution times reported later in this section, we show only transitions between even occupancy states; those for odd occupancy states would have similar behavior.

\begin{figure}[htbp]
    \centering
    \begin{adjustbox}{minipage=18cm, center}
    \begin{subfigure}{\textwidth}
        \centering
        \includegraphics[width=\textwidth, height=0.35\textheight, keepaspectratio, trim=0cm 0cm 0cm 0cm, clip=true]{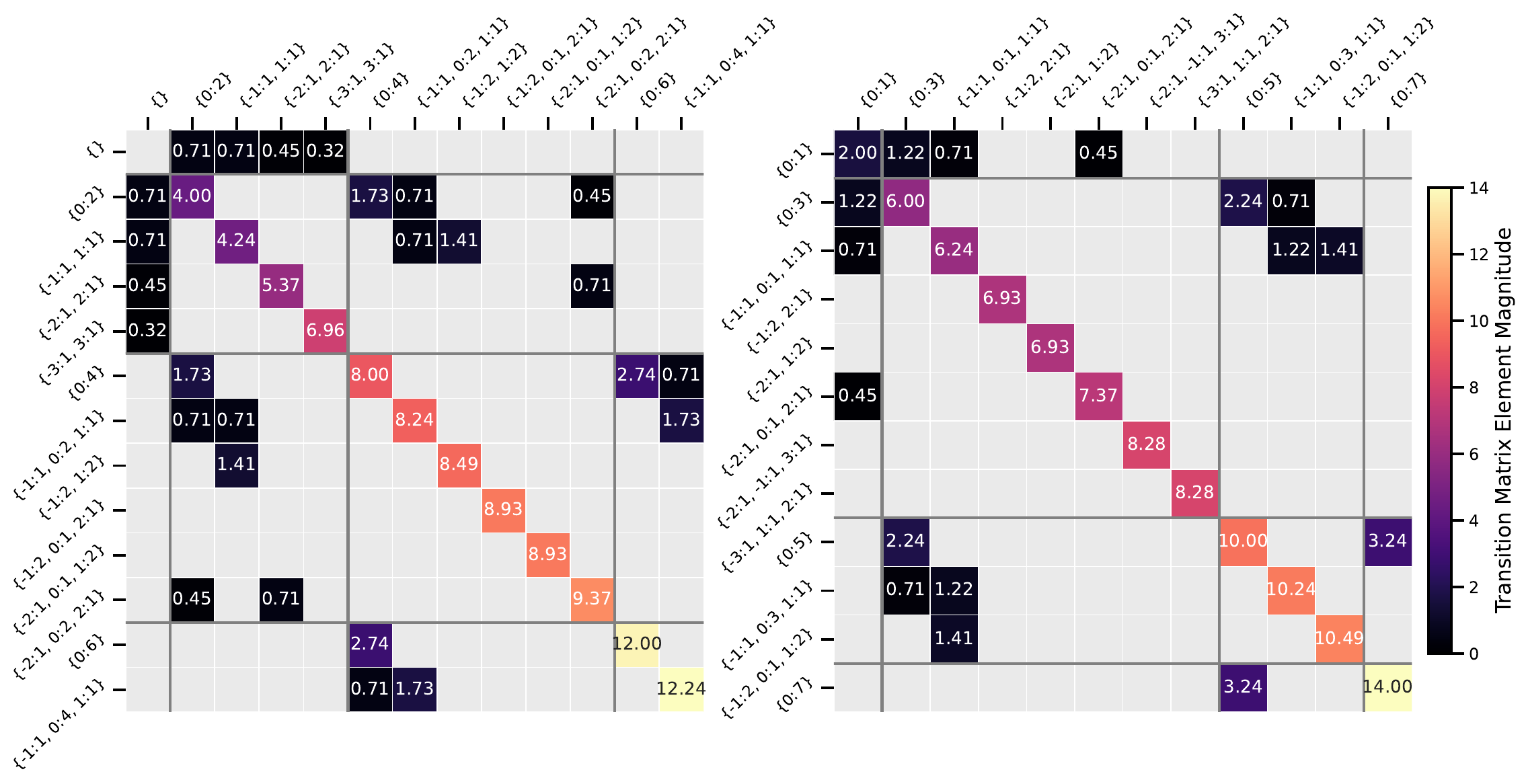} 
        \caption{2D \(\phi^2\) Transition matrix amplitudes, with \(m_V=1.0\).}
        \label{fig:phi2transition}
    \end{subfigure}
    \vspace{0.5cm}
    \begin{subfigure}{\textwidth}
        \centering
        \includegraphics[width=\textwidth, height=0.35\textheight, keepaspectratio, trim=0cm 0cm 0cm 0cm, clip=true]{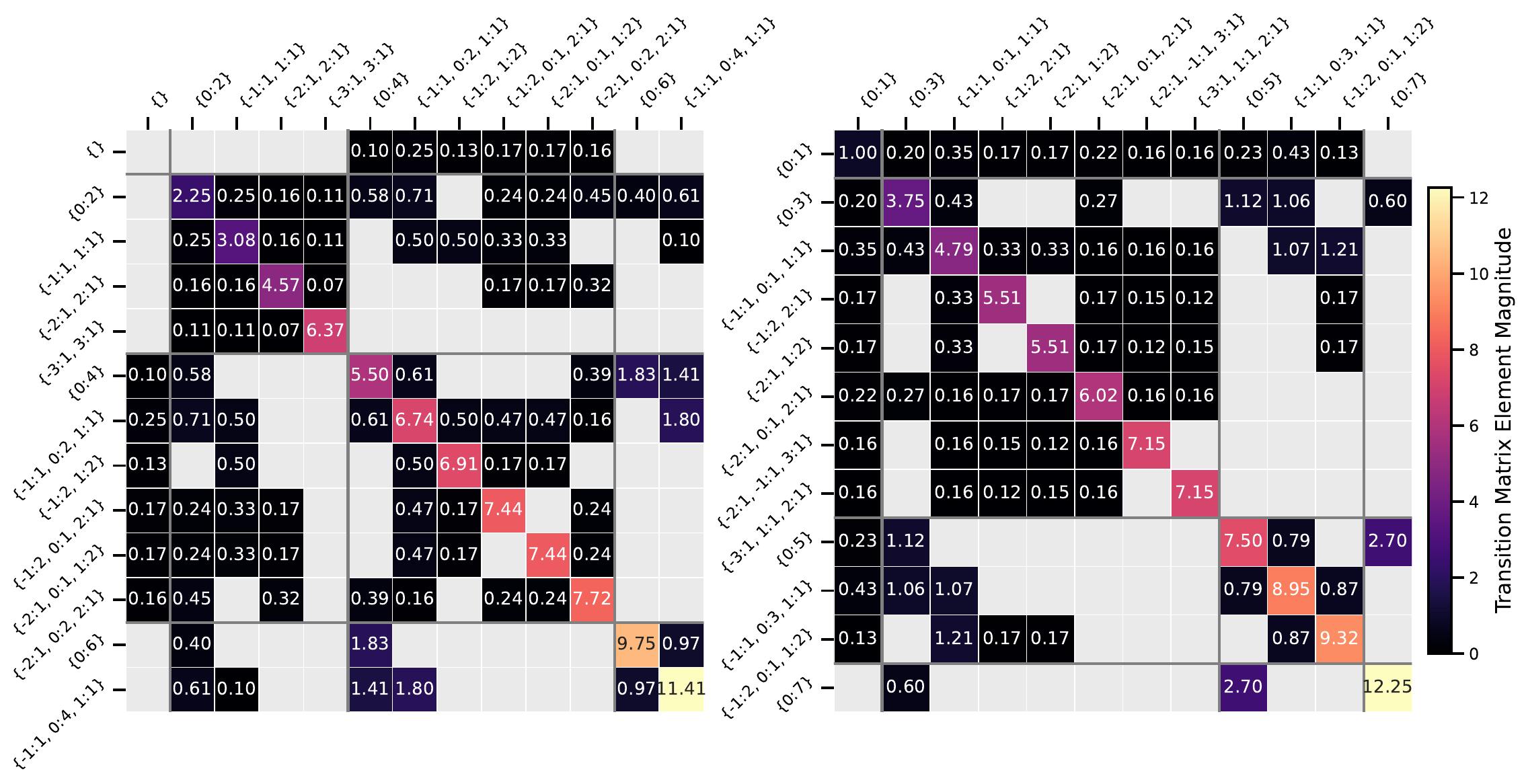} 
        \caption{2D \(\phi^4\) Transition matrix amplitudes, with \(\lambda = 4\pi\).}
        \label{fig:phi4transition}
    \end{subfigure}
    \end{adjustbox}
    \caption{Transition matrix element magnitudes constructed with the parameters \(R=1.0,\, m_Q=1.0,\) and \(E_\text{max}=7\). Further couplings are specified in the individual sub-figure labels. \cref{fig:phi2transition} mirrors \cref{fig:adjacency}, but including now the transition amplitudes and \cref{fig:phi4transition} is the equivalent plot for 2D \(\phi^4\). In this figure, gray boxes mean the transition is forbidden. Note the \(\{\}\rightarrow\{\}\) transition has zero amplitude and is therefore shown in grey here, but is yellow in \cref{fig:adjacency} due to the methodology applied in its construction detailed in \cref{sec:mat-con}.}
    \label{fig:transitionampl}
\end{figure}

Our algorithm is described in pseudocode in \cref{alg:matrix-construction} for the  scenarios \(H_\text{eff}=H_0 + H_2\) (referred to as ``2D \(\phi^2\)") and \(H_\text{eff}=H_0 + H_4\) (referred to as ``2D \(\phi^4\)"). These involve respectively two or four creation and/or annihilation operations which must be normal ordered. \cref{fig:transitionampl} shows example transition matrices for both 2D $\phi^2$ and 2D $\phi^4$ scenarios. The basis has been sorted identically to \cref{fig:adjacency}. Again, a specific implementation of this algorithm in Python can be found in the \href{https://github.com/miarobin/FasterHT}{GitHub repository \faGithub}. 

\begin{algorithm}
\fontsize{10pt}{11pt}\selectfont
\caption{\textbf{Matrix Construction Algorithm.} \newline The required Effective Hamiltonian inputs are \(m_Q,\,R\) and \( E_\text{max}\), as well as \( m_V\) in the case of 2D \(\phi^2\) or \(\lambda\) in the case of 2D \(\phi^4\). The respective transition matrix is returned.}\label{alg:matrix-construction}
\begin{algorithmic}
\PreComputation
\State \mycomment{Set minimum number of annihilations and maximum number of operations.}
\State \textbf{If }\(\phi^2\)\textbf{:} \(i_\text{min}=1\) and \(i_\text{max}=2\); \textbf{else if} \(\phi^4\)\textbf{:} \(i_\text{min}=2\) and \(i_\text{max}=4\)
\Statex \vspace{-0.85em}
\ConstructMatrix
\State \mycomment{This for-loop can be completed in parallel as operations on each state occur independently.}
\For{\texttt{initial\_state}, \texttt{ix}  in \texttt{basis}}

\EndFor
\State \mycomment{(optional) reduce basis to retain only one in $\mathbb{Z}_2$-antisymmetric pairs.}
\For{\texttt{initial\_state}, \texttt{ix} in \texttt{Z2\_reduced\_basis}}
\State \mycomment{Add state energy to diagonal (\(H_0\) operation). (optional) also to $\mathbb{Z}_2$-antisymmetric pairs.}
\State \texttt{transition\_matrix}\,[\texttt{ix},\,\texttt{ix}] = \texttt{energy}\,(\texttt{initial\_state})

\Statex \vspace{-0.85em}
\State\mycomment{Perform \(i=i_\text{min},...,i_\text{max}\) annihilation operations.}
    \For{$\ell_{1},n_{\ell_1}$ in \texttt{initial\_state}}
    \State \texttt{final\_state}\,=\,\texttt{initial\_state}\,[$n_{\ell_1}\rightarrow n_{\ell_1} - 1$]
    \Statex \vspace{-0.85em}
        \For{$\ell_{2}\leq \ell_1, n_{\ell_2}$ in \texttt{final\_state}}
        \SetState \texttt{final\_state}\,[$n_{\ell_2}\rightarrow n_{\ell_2} - 1$]
        \State ...
        \For{$\ell_{i}\leq ...\leq \ell_2 \leq \ell_1, n_{\ell_i}$ in \texttt{final\_state}}
        \SetState \texttt{final\_state}\,[$n_{\ell_i}\rightarrow n_{\ell_i} - 1$]
        \Statex \vspace{-0.85em}
        \State \mycomment{Pre-empt energy ordering of initial and final states, and momentum conservation.}
        \If{\(i=i_\text{max}\)} \(\ell_{i_\text{max}} = - (\ell_1+...+\ell_{i_\text{max}-1})\) \EndIf
        \If{\(i=i_\text{max}-1\)} \(\ell_{i_\text{max}} = + (\ell_1+...+\ell_{i_\text{max}-1})\) \EndIf
        \State \textbf{If }\(\phi^4\) \textbf{and} \(i=2\) \textbf{then for}
            \(k=1,...,\text{floor}[(\ell_1-\ell_2)/2]\) \textbf{do}
            \State\qquad \(\ell_3 = \ell_1 - k\);\, \(\ell_4 = \ell_2 + k\)
            \Statex \vspace{-0.85em}
            \State \mycomment{Perform (all k) creation operations.}
            \For{\(j=i+1,...,(i_\text{max}-i)\)}
            \SetState \texttt{final\_state}\,[$n_{\ell_j}\rightarrow n_{\ell_j} + 1$]
            \EndFor
            \Statex \vspace{-0.85em}
            \State \mycomment{Calculate transition amplitude with Table~\ref{tab:H2matrixelements} or Table~\ref{tab:H4matrixelements}.}
            \State \texttt{entry} = \texttt{transition\_amplitude}\,(\texttt{initial\_state,\,final\_state})
            \Statex \vspace{-0.85em}
            \State \mycomment{Fill transition matrix.}
            \State \texttt{iy} = \texttt{hashable\_basis}[\texttt{final\_state}]; \texttt{transition\_matrix}\,[\texttt{ix},\,\texttt{iy}] = \texttt{entry}
            \Statex \vspace{-0.85em}
            \State \mycomment{Take transpose of off-diagonal elements.}
            \State \texttt{transition\_matrix}\,[\texttt{iy},\,\texttt{ix}] = \texttt{entry}
            \Statex \vspace{-0.85em}
            \State \mycomment{(optional) If using \(\mathbb{Z}_2\)-reduced basis, add required elements.}
            \State \texttt{ix\_Z2}, \texttt{iy\_Z2} = \texttt{find\_Z2\_pair\_indices}\,(\texttt{ix}, \texttt{iy},\texttt{ 
            Z2\_reduced\_basis}) 
            \State \texttt{transition\_matrix}\,[\texttt{ix\_Z2}, \texttt{iy\_Z2}] = \texttt{entry}
            \Statex \vspace{-0.85em}
\Revert \texttt{final\_state} to beginning of for loop.
\State ...
\EndFor
\Revert \texttt{final\_state} to beginning of for loop.
\EndFor
\Revert \texttt{final\_state} to beginning of for loop.
\EndFor
\EndFor
\State
\Return \texttt{transition\_matrix}
\end{algorithmic}
\end{algorithm}

\begin{figure}[htbp]
    \centering

    \begin{subfigure}{\textwidth}
        \centering
        \includegraphics[width=.66\textwidth, trim=0cm 0cm 0cm 0cm, clip=true]{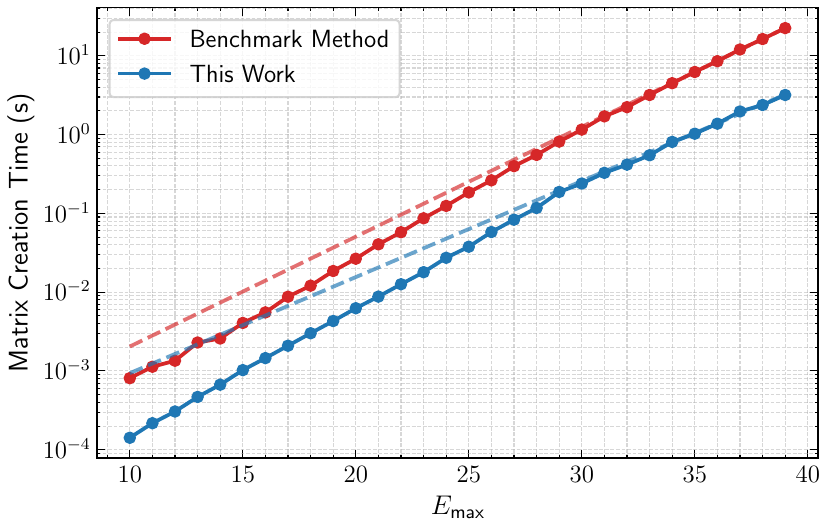} 
        \caption{2D \(\phi^2\) transition matrix construction, run with \(m_V=1.0\).  A linear fit for \(E_\text{max}\in[25,40)\) to the benchmark method has the form \(\log(\tfrac{t}{1s})=0.14\,(E_\text{max})-4.08\), and to the method from this work from \(E_\text{max}\in[25,40)\) the form \(\log(\tfrac{t}{1s})=0.12\,(E_\text{max})-4.25\), where \(t\) is the transition matrix construction time.}
        \label{fig:phi2matrix-time}
    \end{subfigure}
    \hfill
    \begin{subfigure}{\textwidth}
        \centering
        \includegraphics[width=.66\textwidth, trim=0cm 0cm 0cm 0cm, clip=true]{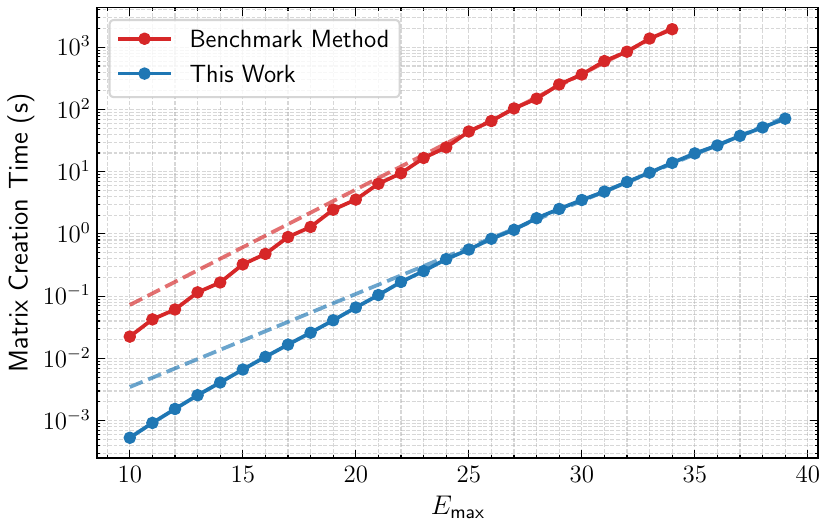} 
        \caption{2D \(\phi^4\) transition matrix construction, run with \(\lambda=4\pi\). A linear fit for \(E_\text{max}\in[25,40)\) to the benchmark method has the form \(\log(\tfrac{t}{1s})=0.19\,(E_\text{max})-2.99\), and to the method from this work from \(E_\text{max}\in[25,40)\) the form \(\log(\tfrac{t}{1s})=0.15\,(E_\text{max})-3.95\).}
        \label{fig:phi4matrix-time}
    \end{subfigure}

    \caption{Timing comparison between the benchmark, \cite{Cohen:2021erm} (red), and the method described in \cref{alg:matrix-construction} (blue) for creating the transition matrix for a 2D $\phi^2$ theory in \cref{fig:phi2matrix-time}, and $\phi^4$ theory in \cref{fig:phi4matrix-time}. The basis construction time is not included in these results. The parameters used to generate these plots are $m_Q = 1.0$ and $R = 1.0$.  The \(E_\text{max}\) starting values for the linear fits are chosen such that the scaling regimes are reached.}
    \label{fig:matrix-time}
\end{figure}

\cref{fig:phi2matrix-time} shows the timing improvement for the 2D $\phi^2$ matrix construction for the method described in Algorithm~\ref{alg:matrix-construction} compared with the benchmark algorithm from \cite{Cohen:2021erm} for an example parameter combination.  \cref{fig:phi4matrix-time} shows the same plot for the case of 2D \(\phi^4\). We note that the matrix construction takes significantly more time than the basis construction. Crucially, the computational time for the method described in this work still scales exponentially, but is nonetheless significantly improved compared to the benchmark method from \cite{Cohen:2021erm}.

The benchmark algorithm proceeds by generating lists of all possible creation and annihilation operators (aside from those related by matrix transpose) which conserve momentum. These operators are then applied to each state in the basis. If the operation succeeds without annihilating the state, it is included in the transition matrix. 

\cref{fig:timingsummary} is the combined timing of the basis and matrix construction as described above, in comparison to \cite{Cohen:2021erm}, for even occupancy states.  Some computational time is required to split the basis into even and odd occupancy states, which can be done at the same time as the creation of the hashable basis if required. This timing cost is also included in the combined construction timing, although it is a small contribution in comparison to the other components. 

With the modifications made to both steps as described in this section, we see a significant decrease in computation time to construct the matrix. Even problems with a reasonably large basis size, here showed up to \(\mathcal{O} (10^6)\), are still tractable classically within a short computation time-frame. For example, in the 2D \(\phi^4\)-case and for such basis sizes, the computational time following this method was of the order \(\sim 10^2\) s, rather than the benchmark algorithm which would be projected to be \(\sim 10^5\) s, i.e., $\sim 1$ day.

Additionally shown in \cref{fig:timingsummary} is the time to calculate the lowest five eigenvalues of the transition matrix (green line). This is completed with the \texttt{scipy.sparse.linalg.eigs} routine as a baseline.\footnote{The \texttt{scipy.sparse.linalg.eigs} also employs Krylov subspace techniques. The quantum Krylov method considered in the following section differs in how the Krylov basis is generated and represented, as the method aims for implementation on quantum hardware.} We see that, while previously the construction of the basis and the transition matrix was the time-limiting operation by a large margin. However, using the methods described in this work the construction of the transition matrix is now comparable in operation time to computation of the five lowest eigenvalues.

We see that for the case of 2D \(\phi^4\), the matrix construction scales similarly with \(E_\text{max}\) to the computation of the low-lying eigenvalues which was not previously the case. Therefore, additionally included in \cref{fig:phi4summary} is the timing for the operation to compute the transition matrix if one parallelizes the loop over states in the basis. This is detailed in Algorithm~\ref{alg:matrix-construction}. We only implement this parallelization (using 4 CPU cores) for the case of 2D \(\phi^4\) as a demonstration.

For smaller values of \(E_\text{max}\), the overhead to split into multiple processes is too large to provide a computational advantage. This is because the basis must be copied to each worker. However, for larger \(E_\text{max}\) values, parallelization does provide a speed-up.  We expect the limiting computational scaling to be similar for matrix construction and calculation of the low-lying eigenvalues. Hence if using the parallelized code, we expect computing the low-lying eigenvalues would be the limiting computational step for large \(E_\text{max}\).
This motivates us to explore quantum-inspired solutions for matrix diagonalization in the following section.

\section{Matrix Diagonalization}
\label{sec:diag}

The truncation constructed in Sec.~\ref{sec:mat-con} yields a finite Hermitian matrix $H$ where the size depends on a given truncation cutoff $E_\text{max}$.
The observables of interest are the vacuum energy, $E_0$, and the low-lying spectral energies, $E_n$ (for small $n$), as well as the energy gaps between these lowest eigenvalues.
The present section concerns their numerical extraction, focusing on quantum computing methods. 

\subsection{Direct Diagonalization}

For a matrix of modest size, the eigenvalues may be obtained directly, for instance by Householder reduction~\cite{golub13} to tridiagonal form followed by the QR algorithm \cite{10.1093/comjnl/4.3.265, 10.1093/comjnl/4.4.332, KUBLANOVSKAYA1962637} or divide-and-conquer, as implemented in standard libraries, e.g.~NumPy~\cite{harris2020array}.
Specifically, we are using NumPy's \verb!linalg.eigvalsh! and \verb!linalg.eigh! functions for exact diagonalization, which internally use LAPACK routines~\cite{doi:10.1137/1.9780898719604}.
This returns the entire spectrum at a cost of $\mathcal{O}(N^3)$ arithmetic operations and $\mathcal{O}(N^2)$ storage.
For the range of $E_\text{max}$ values considered in this work, $N$ will remain small enough so that direct diagonalization is possible, and we use it to obtain reference eigenvalues to machine precision.
These supply the ground truth against which the quantum Krylov method of the following subsection is benchmarked.

\subsection{Quantum Krylov Method}
The principal limitation of Hamiltonian truncation is the exponential growth of the truncated basis with the cutoff $E_\text{max}$ (Fig.~\ref{fig:basis_vs_emax}), which limits the method largely to low-dimensional theories and, even there, caps the accessible $E_\text{max}$.
Quantum computers offer a natural route around this bottleneck, since the dimension of the available Hilbert space grows exponentially with the number of qubits.
First steps toward realizing Hamiltonian truncation on quantum hardware have already been taken~\cite{Ingoldby:2024fcy, Ingoldby:2025bdb}.
Several families of quantum algorithms target the low-lying spectrum of a Hamiltonian.
Quantum phase estimation~\cite{Kitaev:1995qy, Nielsen_Chuang_2010} extracts eigenvalues to, in principle, arbitrary precision, but requires coherent evolution over long times and is therefore a tool for the fault-tolerant era.
Variational quantum eigensolvers~\cite{peruzzo2014variational} are tailored to near-term devices through a hybrid quantum-classical loop, but rely on a problem-specific ansatz and a nonconvex classical optimization that is difficult to control.
Quantum Krylov methods~\cite{Parrish:2019ruc, doi:10.1021/acs.jctc.9b01125, PRXQuantum.3.020323,Motta:2019yya} occupy an intermediate position: they require only moderate circuit depth and replace the variational optimization by a small, well-understood linear-algebra problem.

The basic idea is inherited from classical Krylov methods.
Given a reference state $\vert\phi_0\rangle$, the classical Krylov (Lanczos) approach builds the subspace
\begin{equation}
\mathcal{K}_m = \mathrm{span}\{\vert\phi_0\rangle, H\vert\phi_0\rangle, \dots, H^{m-1}\vert\phi_0\rangle\}\,,
\end{equation}
projects $H$ onto it, and diagonalizes the resulting (much smaller) $m\times m$ matrix.
The extremal eigenvalues converge rapidly in $m$.
Such methods have themselves been applied to truncated-spectrum approaches~\cite{PhysRevD.109.045016}.
On a quantum computer, however, the powers $H^k$ are not directly accessible.
The quantum Krylov method instead generates the subspace by real-time evolution
\begin{equation}
  \mathcal{K}_m = \mathrm{span}\bigl\{\,\vert\phi_0\rangle,\ e^{-i\Delta\tau H}\vert\phi_0\rangle,\
  \dots,\ e^{-i(m-1)\Delta\tau H}\vert\phi_0\rangle\,\bigr\},
\label{eq:K_m_Q}  
\end{equation}
for a time step $\Delta\tau$.
The advantage is that real-time evolution $e^{-i\Delta\tau H}$ is exactly the operation a quantum device implements natively, whereas the powers $H^k$ are not.
Because the basis states in (\ref{eq:K_m_Q}) are not orthogonal, projecting $H$ onto $\mathcal{K}_m$ yields a \emph{generalized} eigenvalue problem 
\begin{equation}
\mathbf{H}\,\vec{c} = E\,\mathbf{S}\,\vec{c}\,,   
\label{eq:small_problem}
\end{equation}
where the projected Hamiltonian $\mathbf{H}$ (shape $m\times m$) and the overlap matrix $\mathbf{S}$ would be assembled from quantities measured on the device and the small problem represented by eq.~(\ref{eq:small_problem}) is then solved classically.
The disadvantage is that the overlap matrix $\mathbf{S}$ becomes ill-conditioned as the basis grows, which must be controlled by regularization~\cite{Epperly:2021ugt, Kirby2024analysisofquantum}.
In the following, we denote the dimension of the Krylov subspace by $N_K$ instead of the generic index $m$.

\paragraph{Setup for the Quantum Krylov method:}
Fix a reference state $\vert\phi_0\rangle$ and a time step $\Delta\tau$, and define the non-orthogonal Krylov basis
\begin{equation}
  \vert\phi_j\rangle \equiv e^{-i j\Delta\tau H}\vert\phi_0\rangle , \qquad j = 0, 1, \dots, N_K-1 .
\end{equation}
Projecting the eigenvalue problem onto $\mathcal{K}_{N_K} = \mathrm{span}\{\vert\phi_j\rangle\}$
requires the projected Hamiltonian~$\mathbf{H}$ and the overlap (Gram) matrix~$\mathbf{S}$,
\begin{align}
  \mathbf{H}_{jk} &= \langle\phi_j\vert H \vert\phi_k\rangle
    = \langle\phi_0\vert\, H\, e^{-i(k-j)\Delta\tau H}\vert\phi_0\rangle , \\
  \mathbf{S}_{jk} &= \langle\phi_j\vert\phi_k\rangle
    = \langle\phi_0\vert\, e^{-i(k-j)\Delta\tau H}\vert\phi_0\rangle ,
\end{align}
where we used $[H, e^{-i k\Delta\tau H}] = 0$. Both matrices are Hermitian and depend on the indices only through the difference $k-j$, i.e.\ they are Toeplitz.
The entire problem is therefore fixed by the real-time autocorrelation function $\langle\phi_0\vert e^{-i n\Delta\tau H}\vert\phi_0\rangle$ and the related correlators $\langle\phi_0\vert H e^{-i n\Delta\tau H}\vert\phi_0\rangle$ at the $N_K$ integer multiples $n = 0, \dots, N_K-1$.
On quantum hardware, these are the quantities accessed through standard circuits such as the Hadamard test~\cite{Nielsen_Chuang_2010}.
In the absence of an easy access to a real quantum device, in the present work the Krylov states are instead constructed by exact classical simulation (the truncated Hamiltonian is diagonalized once and $e^{-i j \Delta\tau H}$ is applied in its eigenbasis) and $\mathbf{H}$ and $\mathbf{S}$ are assembled directly from the states rather than from the correlators, so their Toeplitz structure holds up to floating point errors (this feature will not be exploited, but could prove useful in the future).
This emulation serves as a correctness testbed for the method, not as a scalable classical algorithm.

The approximate eigenpairs (eigenvalue+eigenvector) follow from the generalized eigenvalue problem represented by eq.~(\ref{eq:small_problem})
whose ``eigenvalues'' $E_i$, called \textit{Ritz values}, approximate the lowest eigenvalues of the truncated $H$, with corresponding Ritz vectors $\vert\psi_i\rangle = \sum_{k} (\vec c_i)_k \vert\phi_k\rangle$.
Since $\mathcal{K}_{N_K}$ is a subspace of the truncated Hilbert space, the ordered Ritz values are variational upper bounds on the eigenvalues of the truncated Hamiltonian~\cite{parlett1998symmetric, golub13}: by the Courant--Fischer min--max principle, restricting the variational subspace can only raise each ordered eigenvalue, so the Krylov error is one-sided and adds to the truncation error level by level.
This property survives the thresholding regularization, which merely replaces $\mathcal{K}_{N_K}$ by a subspace of itself~\cite{Epperly:2021ugt}. 

\paragraph{Regularization:}
The generalized eigenvalue problem is well posed only while $\mathbf{S}$ is well conditioned.
As the basis grows with larger $N_K$, the real-time states $\vert\phi_j\rangle$ become progressively linearly dependent:
successive vectors differ only through the phases $e^{-i j\Delta\tau E_n}$ on the populated energy eigenstates.
So the eigenvalues of $\mathbf{S}$ decay rapidly and the smallest ones sink below the level at which $\mathbf{H}$ and $\mathbf{S}$ are known.
On hardware that level is set by shot noise; in the classical simulation used here it is machine precision.
Inverting $\mathbf{S}$ naively then amplifies this noise into spurious eigenvalues.
We control this with the standard thresholding (canonical-orthogonalization) procedure~\cite{Epperly:2021ugt, Kirby2024analysisofquantum}.
Diagonalizing the Hermitian, positive-semidefinite overlap matrix
\begin{equation}
  \mathbf{S} = \sum_{a} \sigma_a\, \vec{v}_a \vec{v}_a^{\dagger}\,,
  \qquad \sigma_1 \ge \sigma_2 \ge \dots \ge 0\,,
\end{equation}
we retain only the subspace spanned by eigenvectors with $\sigma_a > \varepsilon\,\sigma_1$ for a fixed relative threshold $\varepsilon$, and solve the eigenvalue problem projected onto it.
The threshold separates the physically meaningful directions from those dominated by noise:
too small and the residual ill-conditioning re-enters, too large and genuine spectral information is discarded.
Ref.~\cite{Kirby2024analysisofquantum} refines this analysis for real-time quantum Krylov methods, proving ground-state energy error bounds linear in the noise level.
Throughout we set $\varepsilon = 10^{-13}$.

\paragraph{Reference state:}
For the reference state $\vert\phi_0\rangle$, we take the free Fock vacuum in the truncated even-sector Hilbert space.
Since $H$ is block-diagonal in $\mathbb{Z}_2$ parity, the Krylov subspace stays within the parity sector of the reference throughout.
We take the even block, of dimension $n_\text{basis}$.
The free vacuum is not an eigenstate of the interacting truncated Hamiltonian, so it retains nonvanishing overlap with the low-lying states, and the extremal Ritz values converge without problem-specific state preparation.
Its weight concentrates on a small number of low, well-separated levels rather than on the dense high-energy part of the spectrum.
This keeps the frequencies that populate $\mathbf{S}$ few and well spaced, which improves the conditioning of $\mathbf{S}$ and lets the thresholding regularization retain a useful subspace at each depth $N_K$.

\paragraph{Time step:}
The method finds the energies by evolving the reference state in time, not by diagonalizing $H$. What it has access to is the signal
\begin{equation}
  g(t) = \langle\phi_0| e^{-iHt} |\phi_0\rangle = \sum_n w_n\, e^{-iE_n t}\,,
  \qquad w_n = \vert\langle n\vert\phi_0\rangle\vert^2\,,
\end{equation}
sampled at integer multiples of $\Delta\tau$.
Each energy $E_n$ shows up as a phase that turns at its own rate, and the weight $w_n$ is how strongly the reference state excites it.
The overlap matrix is this signal at the sample lags, $\mathbf{S}_{jk} = g\big((j-k)\Delta\tau\big)$, and $\mathbf{H}$ is the same sum with each term multiplied by $E_n$.
So the whole calculation rests on two choices: the spacing
$\Delta\tau$ between samples, and the total time $T = N_K\,\Delta\tau$ they span.
Each one controls a different aspect.
\begin{itemize}
    \item
The spacing $\Delta \tau$ controls \textit{aliasing}.
Since every energy enters only as a turning phase, sampling too slowly makes a fast phase look identical to a slow one, the way a wheel filmed at too few frames appears to spin at a different rate or even backwards.
Two energies that land on the same phase at every sample produce the same data and cannot be discriminated.
Only the spread of the spectrum matters here, not its overall size, because adding a constant to every energy turns all the samples by the same amount, which then cancels out.\footnote{Under $H \to H + c\,I$ every Krylov vector picks up the same phase $e^{-ic\,j\Delta\tau}$, so $\mathbf{S} \to D^\dagger \mathbf{S} D$ and $\mathbf{H} \to D^\dagger(\mathbf{H} + c\,\mathbf{S})D$ with $D$ unitary and diagonal. Every Ritz value shifts by $c$ and nothing else changes.}
The scale that sets the limit is therefore the range $r\equiv E_n^{\rm (max)} - E_n^{\rm (min)}$.
Keeping every energy on its own phase requires $\Delta\tau\,r < 2\pi$~\cite{PRXQuantum.3.020323, Kirby2024analysisofquantum}, the sampling limit familiar from signal processing.

This limit is pessimistic, however, because it treats every energy as if the reference excited it equally, which is not the case.
An energy enters the signal only through its weight $w_n$,
and our reference is the vacuum, which has weight on just a few low, well separated levels.
The high energies that a coarser step would misrepresent as lower ones carry almost no weight, so confusing them costs nothing as long as the reference has negligible weight more than $2\pi/\Delta\tau$ above the lowest energy.
We can therefore step past the naive limit, which is not only safe but useful.
A longer step gives a longer total time $T$ at fixed depth, and that is exactly what closely spaced levels need in order to separate.
We set
\begin{equation}
  \Delta\tau = f\,\frac{2\pi}{r}\,, \qquad f = 1.6\,,
  \label{eq:delta_tau}
\end{equation}
recomputed at each cutoff.
Overshooting (too large of a $\Delta\tau$) has a clear signature: the overlap matrix stays full rank at every depth, so the thresholding retains all $N_K$ directions, and the accuracy per Krylov state drops.
We chose $f$ through a small numerical sweep once, checking that the overlap eigenvalues $\sigma_a$ still decay rapidly, and then kept it fixed.
\item
The total time $T = N_K \Delta\tau$ controls resolution.
Two levels split by $\delta E$ turn at almost the same rate, so at short times they look the same and pull apart only once their phases have drifted.
That takes a total time 
\begin{equation}
T = N_K\,\Delta\tau \gtrsim 1/\delta E.
\label{eq:condition}
\end{equation}
This is a factor of $2\pi$ easier than the same statement for an ordinary discrete Fourier transform, because the recovered energies are not pinned to a fixed frequency grid~\cite{Kirby2024analysisofquantum}.
We checked that only $T$ matters, not how it is divided, by reaching the same $T\,\delta E$ with different $\Delta\tau$ and $N_K$. The condition (\ref{eq:condition}) is enough only when the two levels carry comparable weight, since the weaker one must also stand above the overlap cutoff.
Putting the two limits~(\ref{eq:delta_tau}) and~(\ref{eq:condition}) together gives the depth needed to resolve a splitting,
\begin{equation}
  N_K \gtrsim \frac{1}{\delta E\,\Delta\tau} = \frac{r}{2\pi f\,\delta E}\,.
\end{equation}
The range $r$  grows as a power of the truncation scale $E_{\rm max}$,
while the size of the basis grows faster than any power.
The depth needed to resolve a fixed splitting therefore grows far more slowly than the space it searches, 
which we quantify in Sec.~\ref{sec:phi4_results} (Fig.~\ref{fig:break_even}) below.
\end{itemize}

Setting the step needs $r$, but only as an upper bound. Any upper bound $r_{\rm ub}$ implies $r_{\mathrm{ub}} \ge r$, which still keeps $\Delta\tau$ within the sampling limit.
It only makes the step a little smaller, and since $N_K \propto r$, using a bound in place of the true range raises the depth by the ratio $r_{\mathrm{ub}}/r$ and nothing worse.
This is what makes the recipe work on hardware.
Finding $r$ exactly would mean diagonalizing $H$, the very job the method is meant to avoid, but a bound needs no diagonalization.
Gershgorin's theorem~\cite{golub13} gives one straight from the matrix elements, at a cost that scales with the number of nonzero entries.
A tighter estimate is available while running: the weights $w_n$ can be read from the Fourier transform of the measured overlaps, which are collected anyway, and they show the range of energies which the reference actually populates.
In the classical emulation here, we use the exact $r$, and treat the bounded step as a separate resource estimate.

\section{Quantum Diagonalization Results}
\label{sec:results}
\subsection{2D \texorpdfstring{$\phi^2$}{phi2} Theory}
\label{sec:phi2_results}
\begin{figure}[h]
\centering
\includegraphics[width=\textwidth]{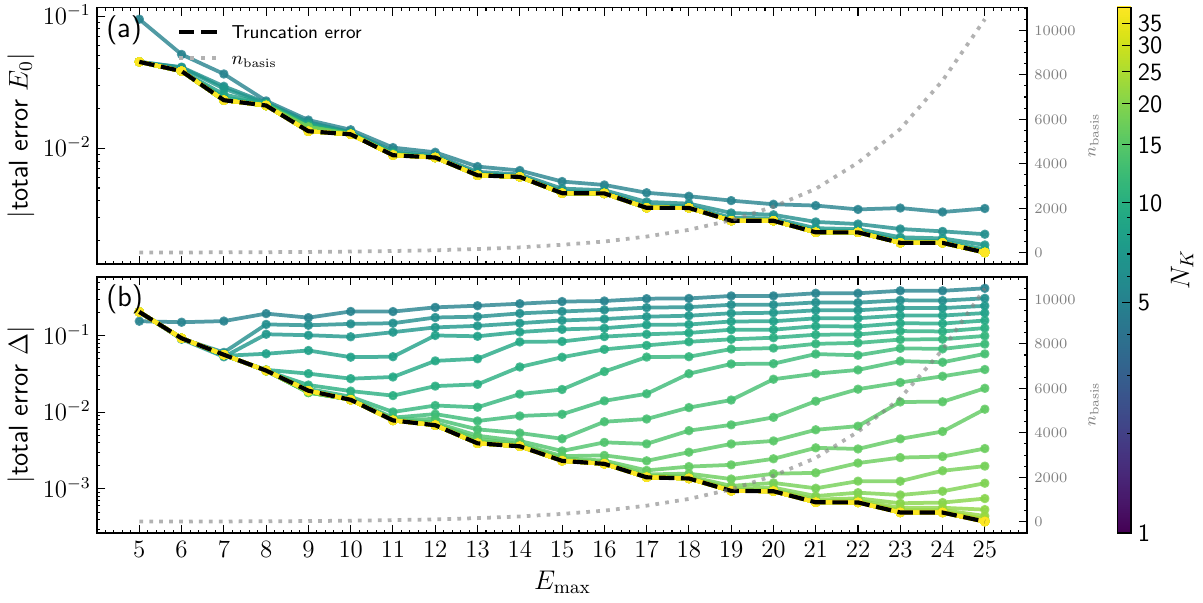} 
\caption{Total error (Krylov $+$ truncation) relative to the exact analytical spectrum of the $\phi^2$ theory in the even sector, as a function of $E_\text{max}$ ($m_Q=1.0$, $m_V=1.0$, $R=1.0$).
The color indicates the number of Krylov states $N_K$.
(a) Error of the ground state $E_0$, (b) error of the gap $\Delta$ between $E_0$ and the first excited state in the even sector.
The black dashed line is the truncation error alone, i.e.\ the deviation of the exactly diagonalized truncated Hamiltonian from the exact analytical result.
The gray dotted line (right-hand axis) shows the basis size $n_\text{basis}$ and thus the size $(n_\text{basis}\times n_\text{basis})$ of the truncated Hamiltonian.}
\label{fig:total_err}
\end{figure}

Because $\phi^2$ theory is exactly solvable, we can separate the two sources of error in the Krylov spectrum: the \emph{truncation error} of the finite-$E_\text{max}$ Hamiltonian relative to the exact analytical result, and the additional \emph{Krylov error} incurred by working in the $N_K$-dimensional subspace rather than diagonalizing the full truncated matrix.
The truncation error (black dashed in Fig.~\ref{fig:total_err}) falls monotonically with $E_\text{max}$, from $\sim\!4\times10^{-2}$ to $\sim\!10^{-3}$ for $E_0$ and from $\sim\!2\times10^{-1}$ to $\sim\!4\times10^{-4}$ for the gap $\Delta$.
The central observation is that, given sufficient depth, the total error saturates this floor: the high-$N_K$ curves lie directly on the truncation line, so the Krylov step reproduces the truncated spectrum to within the truncation error itself and contributes negligibly beyond it.

\begin{figure}[h]
\centering
\includegraphics[width=\textwidth]{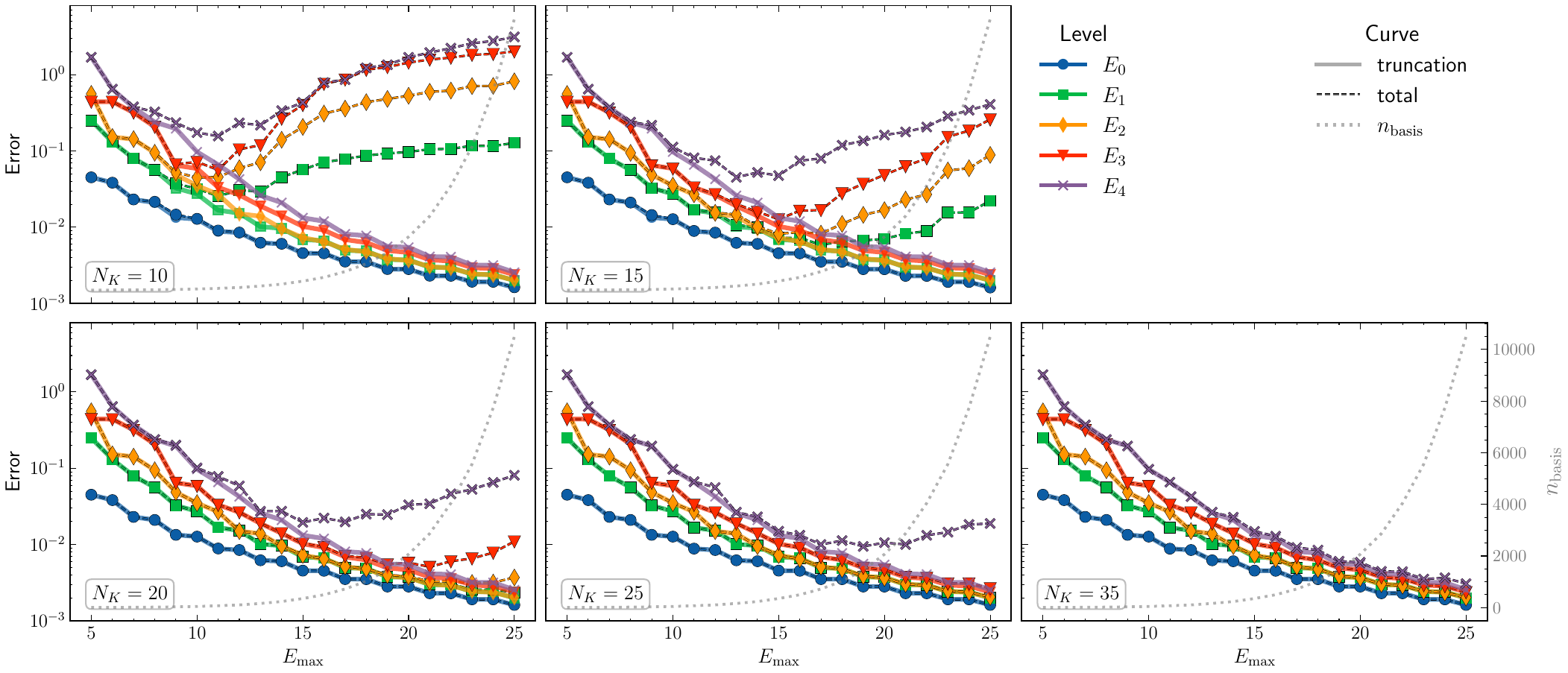}
\caption{Truncation error and total error as a function of $E_\text{max}$ for the $\phi^2$ theory, shown for several Krylov basis sizes $N_K$ (one panel each). In this figure, \(m_Q=1.0,\,R=1.0\) and \(m_V=1.0\).
Colors distinguish the energy levels $E_0$ through $E_4$. 
For each level, the thick solid line is the truncation error and the dashed line is the total error (truncation~$+$~Krylov), evaluated at the corresponding $N_K$.
The dotted gray line (right-hand axis) shows the basis size $n_\text{basis}$ of the truncated Hamiltonian.}
\label{fig:levels_grid}
\end{figure}

\begin{figure}[h]
\centering
\includegraphics[width=.6\textwidth, trim=0cm 0cm 0cm .7cm, clip=true]{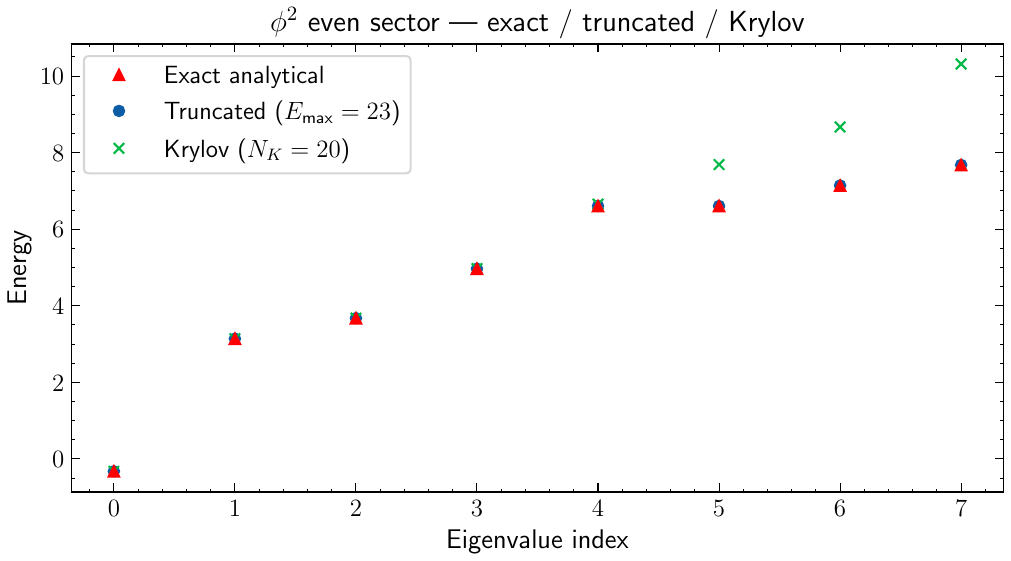}
\caption{Comparison of the lowest even-sector eigenvalues of the $\phi^2$ theory obtained from the exact analytical spectrum, from exact diagonalization of the truncated Hamiltonian with \(m_Q=1.0,\,R=1.0\) and \(m_V=1.0\) at $E_\text{max}=23$, and from the Krylov method at $N_K=20$.
The three agree for the lowest levels.
From the sixth level (index~5) upward the Krylov values are biased upward, as required by the variational bound.
}
\label{fig:evals}
\end{figure}

Convergence proceeds from the bottom of the spectrum upward and level by level.
The ground state converges first and cheapest, while the gap and the higher excited states require larger $N_K$ (Fig.~\ref{fig:total_err}(a) vs.~\ref{fig:total_err}(b), and the per-level panels of Fig.~\ref{fig:levels_grid}):
at small $N_K$, the total error for $E_3$ and $E_4$ departs from the truncation line at some $E_\text{max}$,  climbing steadily once the evolution is too short to resolve the higher levels, consistent with the criterion $T\gtrsim 1/\delta E$.
As $N_K$ grows these departures recede to higher $E_\text{max}$ until, by the largest depth shown ($N_K=35$), all five levels track truncation across the full range.
The snapshot in Fig.~\ref{fig:evals} shows the same behavior at fixed $(E_\text{max},N_K)=(23,20)$: 
the lowest five even-sector levels coincide with the exact and truncated values, while from the sixth level upward the Krylov values are biased upward. 
This one-sidedness is guaranteed by the variational bound and is the hallmark of bottom-up Krylov convergence.

\begin{figure}[h]
\centering
\includegraphics[width=\textwidth, trim=0cm 0cm 0cm 0cm, clip=true]{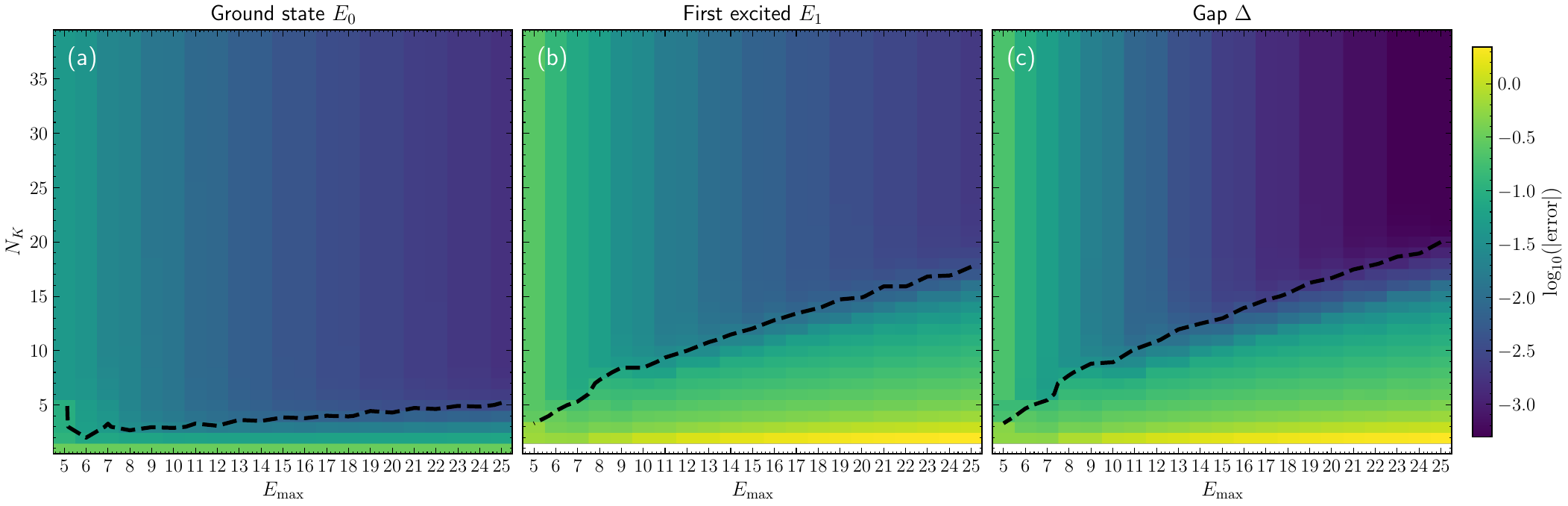}
\caption{Heatmap of the total error (Krylov $+$ truncation) relative to the exact analytical spectrum of the $\phi^2$ theory in the even sector, as a function of $E_\text{max}$ and $N_K$ ($m_Q=1.0$, $m_V=1.0$, $R=1.0$).
         (a) Ground state $E_0$, (b) first excited state of the even sector $E_1$, (c) gap $\Delta = E_1-E_0$.
         The black dashed line indicates where the Krylov error equals the truncation error.
         Above it the result is truncation-limited, below it Krylov-limited.}
  \label{fig:heatmap}
\end{figure}

Finally, the heatmaps of Fig.~\ref{fig:heatmap} quantify the cost.
The dashed contour marks where the Krylov and truncation errors are equal.
Above it, the result is truncation-limited (the Krylov subspace has converged), below it Krylov-limited.
This break-even depth grows with $E_\text{max}$, driven, as discussed above, by the shrinking of $\Delta\tau\propto 1/r$ rather than by any intrinsic degradation of the method.
It rises to $N_K \approx 5$ for $E_0$ and $N_K\approx 18$ for both $E_1$ and the gap at $E_\text{max}=25$,
remaining far below the basis dimension $n_\text{basis}$, which itself grows steeply (gray dotted, Figs.~\ref{fig:total_err} and \ref{fig:levels_grid}).
Reaching truncation-limited accuracy thus requires far fewer Krylov vectors than the size of the truncated Hamiltonian.
Even in this classically tractable testbed, the method compresses the problem substantially.

\subsection{2D \texorpdfstring{$\phi^4$}{phi4} Theory}
\label{sec:phi4_results}
Unlike $\phi^2$, the $\phi^4$ theory is not exactly solvable, so we no longer have a closed-form spectrum against which to estimate the truncation error.
Fortunately, two references survive the loss of the exact solution and take on complementary roles.
The classical diagonalization of the finite-$E_\text{max}$ Hamiltonian remains available at every cutoff we consider, and it is the exact reference for the \emph{Krylov error}.
This error is the deviation of the $N_K$-dimensional Ritz values from the eigenvalues of the full truncated matrix.
It is sharply defined, and it never invokes the extrapolated answer.
The extrapolated spectrum itself is estimated instead by the $E_\text{max}\to\infty$ extrapolation of Appendix~\ref{sec:phi4_extrapolation}.
This follows a power law $\Delta E_n(E_\text{max}) = \Delta E_n^\infty + c_p/E_\text{max}^{\,p}$, with $p$ fixed by dimensional analysis and confirmed with Hamiltonian Truncation Effective Theory (HTET) power counting~\cite{Cohen:2021erm, Demiray:2025zqh}.
This supplies an estimate $E_n^\infty$, and it is the reference for the \emph{truncation error}.

The total error then decomposes exactly as in the solvable case,
\begin{equation}
  \underbrace{E_n^{\text{Kry}}(E_\text{max},N_K) - E_n^\infty}_{\text{total}}
  = \underbrace{E_n^{\text{Kry}}(E_\text{max},N_K)
      - E_n^{\text{trunc}}(E_\text{max})}_{\text{Krylov}}
  + \underbrace{E_n^{\text{trunc}}(E_\text{max}) - E_n^\infty}_{\text{truncation}},
  \label{eq:phi4_error_split}
\end{equation}
the Krylov term compared to the full diagonalization and the truncation term to the extrapolated value.
The only change relative to $\phi^2$ is that $E_n^\infty$ now carries the combined statistical and systematic uncertainty of the fit.
For the excited levels there is an additional uncertainty from the vacuum extrapolation, since the absolute reference is reconstructed as $E_n^\infty = E_0^\infty + \Delta E_n^\infty$.
Unlike the sharp truncation reference of the $\phi^2$ case (Fig.~\ref{fig:total_err}), the $\phi^4$ floor is therefore known only up to the uncertainty of the fit.
The black dashed line in the figures is its best estimate, and the caveats below set how far down it can be trusted.

The Krylov error, by contrast, is untouched by all of this.
It lives entirely inside the truncated theory, compared to the classical diagonalization rather than to the $E_{\rm max}\to\infty$ extrapolation.
It is therefore as sharply defined here as it was for $\phi^2$.

\begin{figure}
    \centering
    \includegraphics[width=\linewidth]{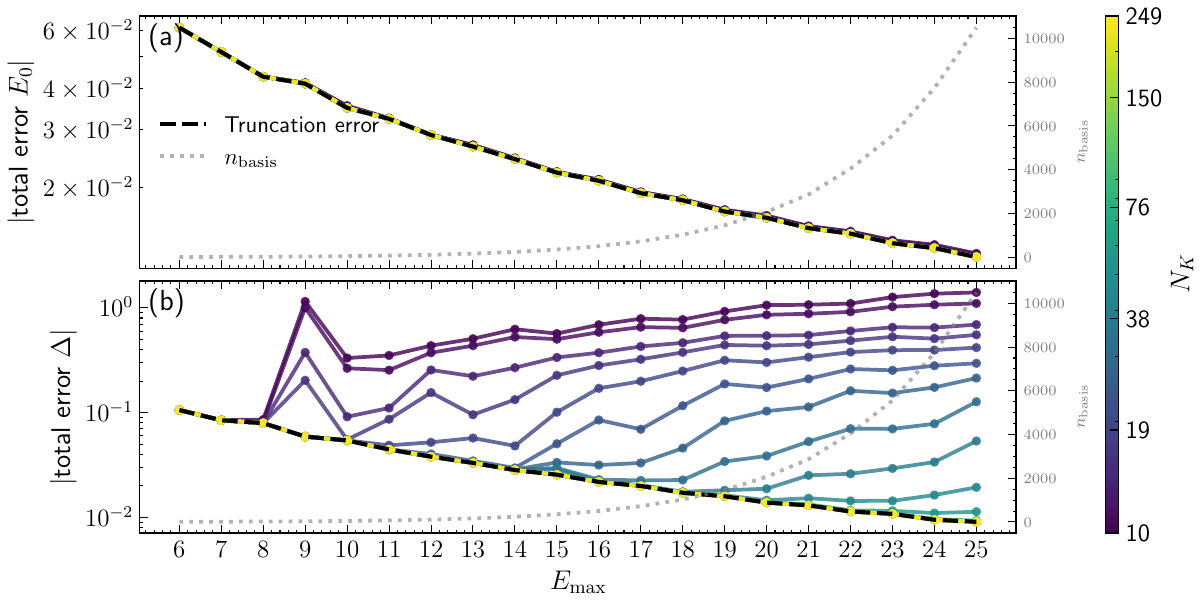}
    \caption{Total error (Krylov $+$ truncation) relative to the extrapolated $E_\text{max}\to\infty$ baseline for the $\phi^4$ theory in the even sector, as a function of $E_\text{max}$ ($m_Q=1.0$, $R=1.0$, $\lambda=4\pi$).
    The color indicates the number of Krylov states $N_K$.
    (a) Error of the ground state $E_0$, (b) error of the gap $\Delta$ between $E_0$ and the first excited state in the even sector.
    The black dashed line is the truncation error alone, i.e.\ the deviation of the exactly diagonalized truncated Hamiltonian from the extrapolated baseline.
    The gray dotted line (right-hand axis) shows the basis size $n_\text{basis}$.
    }
    \label{fig:phi4_total_errors}
\end{figure}

The truncation error (black, dashed line in Fig.~\ref{fig:phi4_total_errors}) again falls monotonically with $E_\text{max}$.
It drops from $\sim\!6\times10^{-2}$ to $\sim\!10^{-2}$ for $E_0$, and from $\sim\!10^{-1}$ to $\sim\!10^{-2}$ for the gap $\Delta$.
Given enough depth, the total error saturates this floor: the high-$N_K$ curves lie directly on the truncation line across the full range of $E_\text{max}$.
Because the Krylov error is measured against the classical diagonalization, this saturation does not rely on the extrapolation, and it holds even though $\phi^4$ has no exact solution.
For the ground state the floor is already reached at the smallest depth shown: every curve with $N_K\ge10$ tracks the truncation line across the full range (Fig.~\ref{fig:phi4_total_errors}(a)).
The gap is more demanding.
At fixed $N_K$ the total error tracks the floor only up to a break-even value of $E_\text{max}$, beyond which the truncated space has grown too large for $N_K$ Krylov vectors to resolve and the Krylov error overtakes the truncation error.
This break-even point moves to smaller $E_\text{max}$ as $N_K$ decreases.
For the smallest depths shown, the gap departs from the floor already near $E_\text{max}\approx9$, and the total error then climbs by about one order of magnitude by $E_\text{max}=25$.

\begin{figure}
    \centering
    \includegraphics[width=\linewidth]{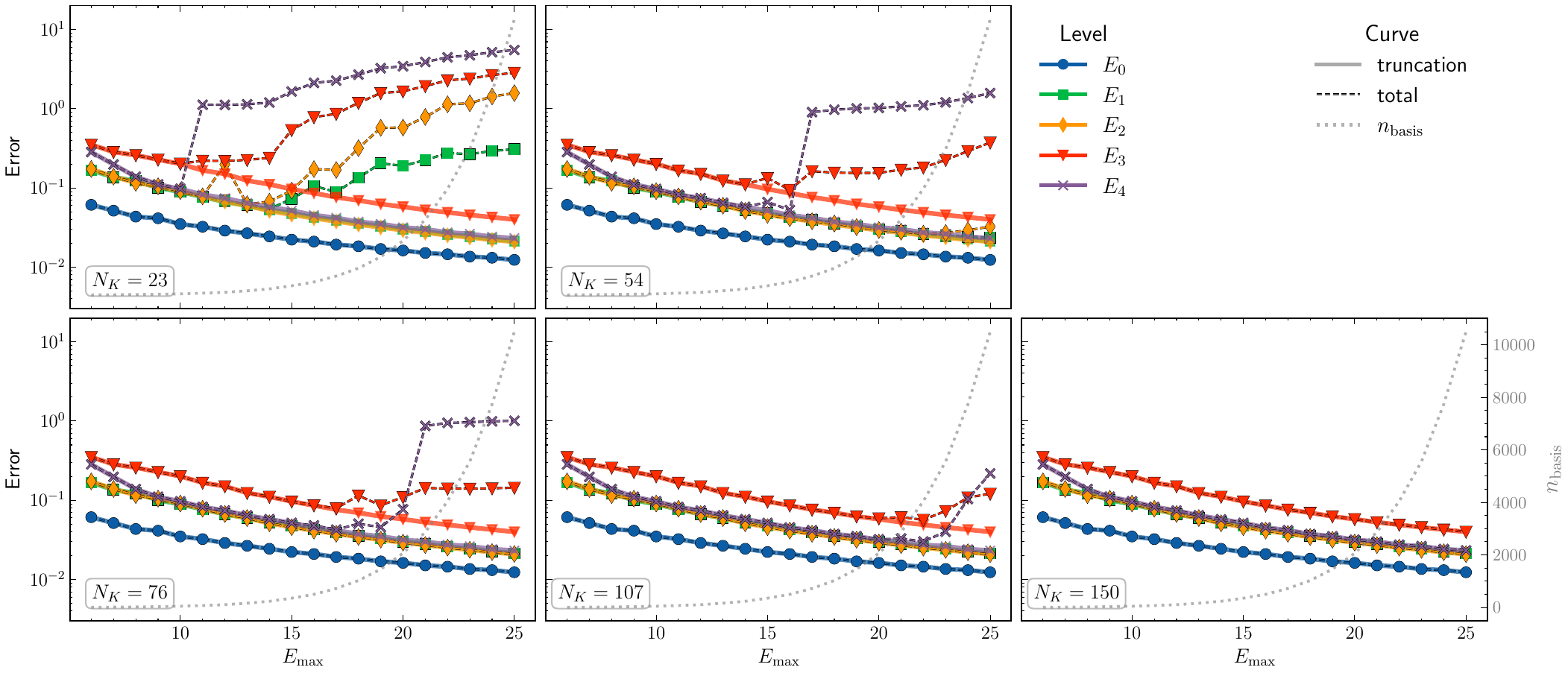}
    \caption{Truncation error and total error as a function of $E_\text{max}$ for the $\phi^4$ theory in the even sector, shown for several Krylov basis sizes $N_K$ (one panel each). The Hamiltonian parameters used in these plots are \(m_Q=1.0,\, R=1.0\) and \(\lambda=4\pi\).
    Colors distinguish the energy levels $E_0$ through $E_4$.
    For each level, the thick solid line is the truncation error and the dashed line is the total error (truncation~$+$~Krylov) at the corresponding $N_K$.
    The dotted gray line (right-hand axis) shows the basis size $n_\text{basis}$.
    Convergence is bottom-up: at small $N_K$ the upper levels detach from their truncation lines at large $E_\text{max}$, and these departures recede as $N_K$ grows.
    }
    \label{fig:phi4_levels_grid}
\end{figure}

\begin{figure}
    \centering
    \includegraphics[width=0.6\linewidth, clip=true, trim=0cm 0cm 0cm .7cm]{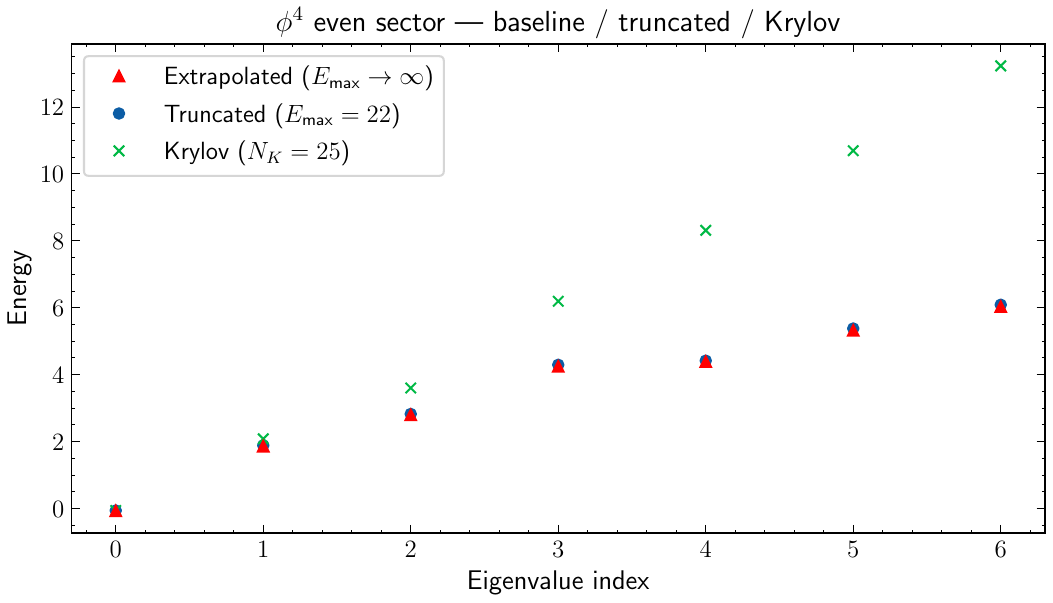}
    \caption{Comparison of the lowest even-sector eigenvalues of the $\phi^4$ theory obtained from the extrapolated baseline ($E_\text{max}\to\infty$), from exact diagonalization of the truncated Hamiltonian with \(m_Q=1.0,\,R=1.0\) and \(\lambda=4\pi\) at $E_\text{max}=22$, and from the Krylov method at $N_K=25$.
    The lowest levels agree.
    From the third level (index~2) upward the Krylov values are biased upward, as required by the variational bound, with the bias growing with the index.
    }
    \label{fig:phi4_evals}
\end{figure}

Convergence proceeds from the bottom of the spectrum upward, level by level~(Fig.~\ref{fig:phi4_levels_grid}).
Larger $N_K$ generally gives smaller Krylov errors.
At $N_K=23$, the upper levels $E_3$ and $E_4$ have already left the truncation line by $E_\text{max}\approx11$ and reach errors of order unity.
As $N_K$ grows, fewer levels deviate and those that do only start to deviate at larger $E_\text{max}$.
By $N_K=107$ all levels except $E_4$ at the largest $E_\text{max}$ track truncation across the entire range and by $N_K=150$ all five levels do.
The snapshot at fixed $(E_\text{max},N_K)=(22,25)$ in Fig.~\ref{fig:phi4_evals} shows the same pattern in the spectrum itself.
The lowest two levels coincide with the truncated and extrapolated values, while from the third level upward the Ritz values are visibly biased upward (as expected from the variational bound), with the bias growing with the index.

\begin{figure}
    \centering
    \includegraphics[width=0.45\linewidth]{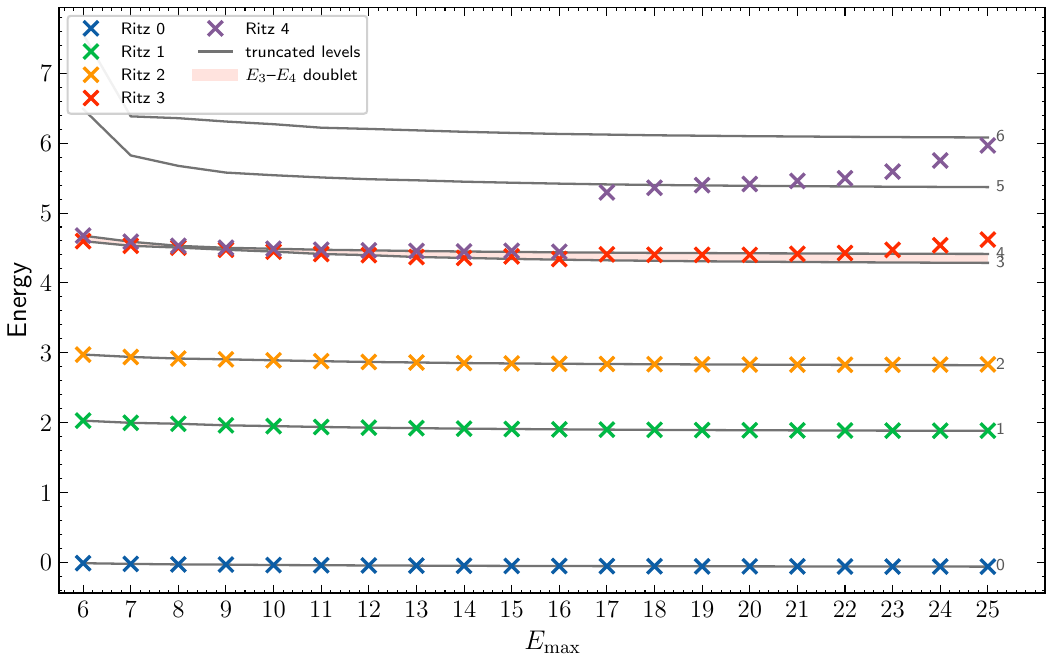}
    \hfill
    \includegraphics[width=0.52\linewidth]{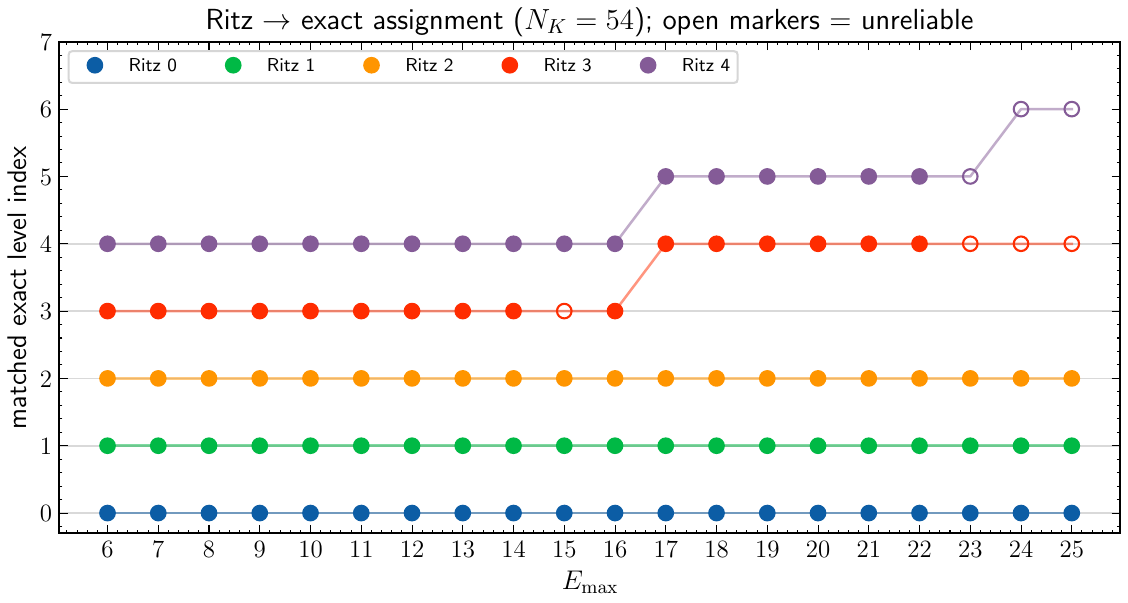}
    \caption{Level assignment at $N_K = 54$ for the $\phi^4$ theory in the even sector. These figures were produced for the sample parameters \(m_Q = 1.0,\, R=1.0\) and \(\lambda=4\pi\). 
      \emph{Left:}~Truncated eigenvalues (gray lines, labeled by index) and Ritz values (colored markers) as functions of $E_\text{max}$.
      The shaded band marks the $E_3$--$E_4$ near-degeneracy.
      \emph{Right:}~For each Ritz index, the matched truncated level index.
      Filled markers are reliable assignments (matching error below 25\% of the local spacing), open markers are ambiguous.
      }
    \label{fig:level_assignment}
\end{figure}

\begin{figure}
    \centering
    \includegraphics[width=\linewidth]{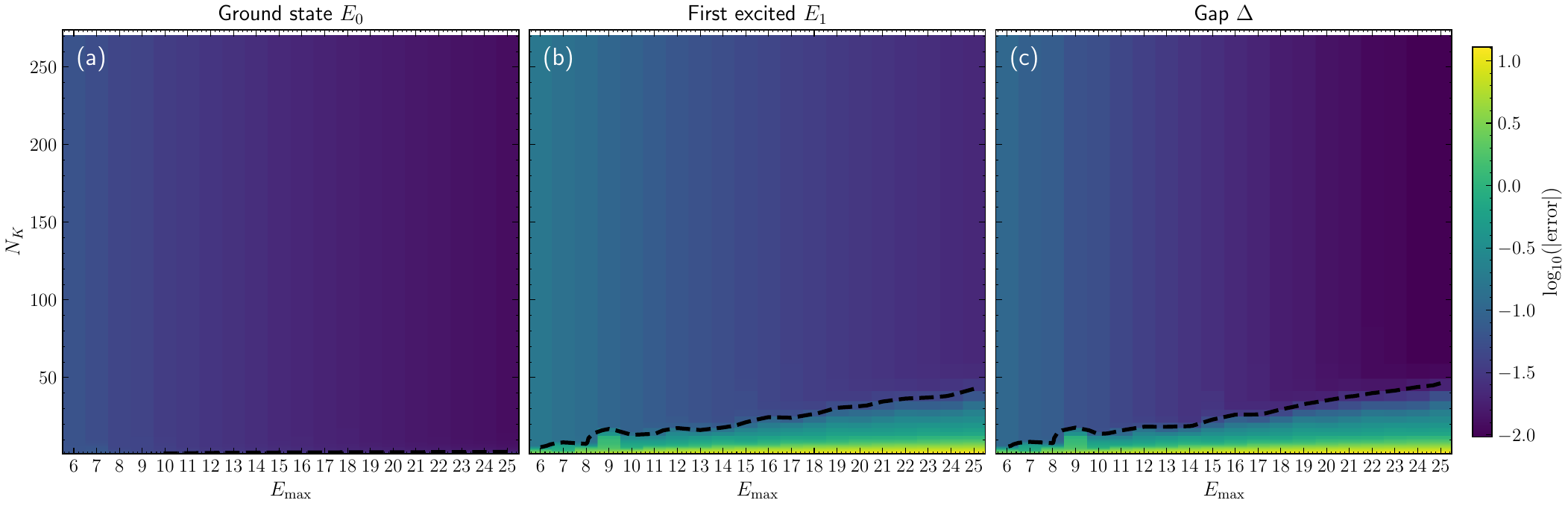}
    \caption{Heatmap of the total error (Krylov $+$ truncation) relative to the extrapolated baseline for the $\phi^4$ theory in the even sector, as a function of $E_\text{max}$ and $N_K$. These figures are produced for the effective Hamiltonian parameters \(m_Q=1.0,\,R=1.0\) and \(\lambda=4\pi\).
    (a) Ground state $E_0$, (b) first excited state $E_1$, (c) gap $\Delta$.
    The black dashed line indicates where the Krylov error equals the truncation error.
    Above it the result is truncation-limited, below it Krylov-limited.
    }
    \label{fig:phi4_heatmap}
\end{figure}

The $E_3$--$E_4$ splitting is roughly an order of magnitude smaller than its neighbors at every $E_\text{max}$ in our range (Fig.~\ref{fig:level_assignment}, left, shaded band).
When the total evolution time $T = N_K\Delta\tau$ drops below $1/(E_4 - E_3)$, the Krylov subspace can no longer resolve the pair: the fourth and fifth Ritz values overshoot both energies and begin tracking higher levels.
Naively comparing the $k$-th Ritz value to the $k$-th truncated level then mistakes the level spacing for the approximation error.
On the right plot of Fig.~\ref{fig:level_assignment}, we therefore assign each Ritz value to the nearest unclaimed truncated level, processing the most unambiguous matches first so that a well-converged low Ritz value claims its level before a poorly converged high one can.
Each match is flagged as reliable (filled marker in Fig.~\ref{fig:level_assignment}, right) when the matching error is below 25\% of the local level spacing, and unreliable otherwise (open markers).
The heatmap contours and error curves of Figs.~\ref{fig:phi4_total_errors}--\ref{fig:phi4_heatmap} involve only $E_0$, $E_1$, and the gap, for which the two lowest Ritz values are unambiguously assigned at every $(E_\text{max}, N_K)$ in our range.
The departures visible for $E_3$ and $E_4$ in Fig.~\ref{fig:phi4_levels_grid} at intermediate $N_K$ are the index-matching artifact described here.

The heatmaps of Fig.~\ref{fig:phi4_heatmap} quantify the cost.
The dashed contour marks where the Krylov and truncation errors are equal.
Above it, the result is \textit{truncation-limited}, below it the result is \textit{Krylov-limited}.
This break-even depth $N_K^*$ grows with $E_\text{max}$.
For the ground state it stays very low, rising only from $N_K\approx1$ to $N_K\approx2$ across the range.
For the first excited state and the gap, it rises from $N_K\approx5$ at the smallest cutoffs to $N_K\approx45$ at $E_\text{max}=25$.
$N_K^*$ is far larger for the excited level and the gap than for the ground state.
The gap inherits the error of both endpoints, so it sits close to the first excited level.
Even so, it stays more than two orders of magnitude below the basis dimension $n_\text{basis}$, which reaches $\approx1.0\times10^4$ at $E_\text{max}=25$ (gray dotted, Figs.~\ref{fig:phi4_total_errors} and \ref{fig:phi4_levels_grid}).

\begin{figure}
    \centering
    \includegraphics[width=\linewidth]{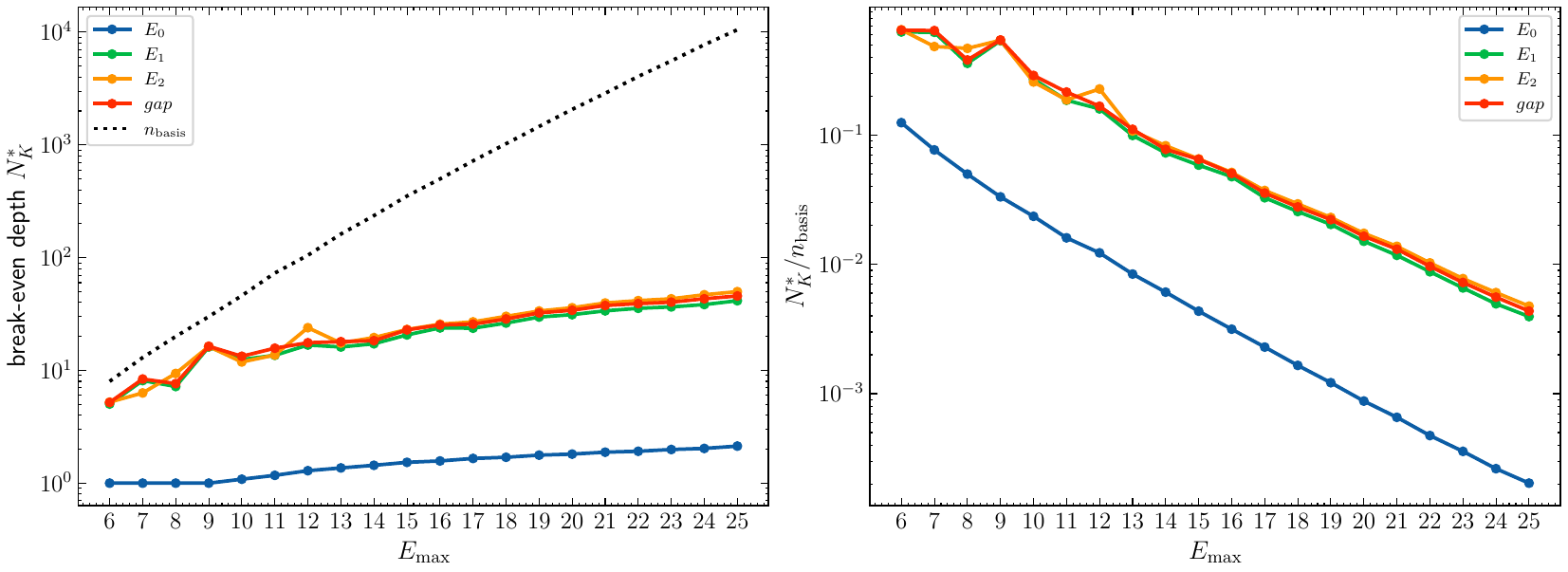}
    \caption{Break-even depth $N_K^*$ for the $\phi^4$ theory in the even sector, as a function of $E_\text{max}$ ($m_Q=1.0$, $R=1.0$, $\lambda=4\pi$).
    At fixed $E_\text{max}$, $N_K^*$ is the smallest Krylov depth at which the Krylov error equals the truncation error.
      (a)~Break-even depth $N_K^*$ as a function of $E_\text{max}$
      for $E_0$, $E_1$, $E_2$, and the gap~$\Delta$.
      The black dotted line shows the basis dimension $n_\text{basis}$.
      (b)~Ratio $N_K^*/n_\text{basis}$.
      The ratio decreases steadily with $E_\text{max}$ for every level,
      so the compression improves as the truncated space grows.}
    \label{fig:break_even}
\end{figure}

The break-even depth $N_K^*$ itself is shown in Fig.~\ref{fig:break_even}.
At fixed $E_\text{max}$, it is the smallest number of Krylov vectors at which the total error reaches the truncation floor, i.e.\ where the Krylov error equals the truncation error.
This is the dashed contour of Fig.~\ref{fig:phi4_heatmap} read as a curve in $E_\text{max}$, here extended to the second even excitation $E_2$.
The left panel plots $N_K^*$ next to the basis dimension $n_\text{basis}$.
The ground state needs one or two Krylov vectors across the whole range.
The excited levels $E_1$ and $E_2$ and the gap track one another and rise from $N_K^*\approx5$ at $E_\text{max}=6$ to $N_K^*\approx45$ at $E_\text{max}=25$.
Over the same range, $n_\text{basis}$ grows from a few tens to $\approx1.0\times10^4$, so at the largest cutoff the break-even depth sits more than two orders of magnitude below the size of the truncated matrix.

The right panel of Fig.~\ref{fig:break_even} states the same result as a fraction.
The ratio $N_K^*/n_\text{basis}$ falls monotonically with $E_\text{max}$ for every level.
For the excited levels and the gap it drops from about one half at $E_\text{max}=6$ to $\sim\!4\times10^{-3}$ at $E_\text{max}=25$, and for the ground state from about $0.1$ to $\sim\!2\times10^{-4}$.
This is the quantitative form of the argument in Sec.~\ref{sec:diag}.
The range $r$ that sets the step grows only as a power of the cutoff, while $n_\text{basis}$ grows faster than any power, so the fraction of the basis needed for truncation-limited accuracy shrinks as the cutoff rises.
The compression is not a fixed factor.
It improves with $E_\text{max}$, which is exactly the regime where a full diagonalization is most costly.

One caveat applies to the extrapolated baseline: the truncation floor is only as sharp as the fit.
The break-even contours are therefore reliable even where the floor they are compared against is not.
Where the truncation error approaches that uncertainty, the black curve measures the quality of the fit rather than a physical truncation error.
This does not affect the Krylov error.
It is compared to the classical diagonalization and stays sharply defined at every $(E_\text{max},N_K)$.
The break-even contours are therefore reliable even where the floor they are compared against is not.

\section{Conclusions}
In this work, we investigated the computational implementation  of Hamiltonian truncation. We focused on three computational steps required to obtain the low-lying spectrum: basis generation, Hamiltonian matrix construction, and matrix diagonalization. We developed an efficient basis-generation algorithm based on integer partitions, introduced a method to reduce the cost of sparse matrix construction, and classically simulated quantum Krylov methods to assess their potential as an alternative approach to diagonalization. 

Our results demonstrate that improvements to the classical stages of the calculation substantially reduce the cost of Hamiltonian truncation. For the benchmark 2D scalar theories considered here, these improvements shift the dominant contribution to the computational cost from matrix construction to matrix diagonalization when standard sparse matrix eigensolvers are employed. No approximation is made at the step of matrix construction itself: every matrix element connecting states within the truncated Hilbert space is retained. This ensures the preservation of separation of scales and makes the resulting matrices compatible with improvement programs based on counterterm corrections. With the optimized matrix construction algorithm presented here, further improvements to the scalability of Hamiltonian truncation will increasingly depend on more efficient approaches to extracting the low-lying spectrum. We expect this shift in the computational bottleneck to be at least as pronounced in fermionic theories, where restricted occupation numbers reduce the size of the truncated Hilbert space for fixed $E_{\rm max}$ while producing less sparse Hamiltonians. 

To address this remaining bottleneck, we studied the quantum Krylov method through classical simulation. We find that accurate low-lying eigenvalues can be obtained using Krylov spaces significantly smaller than the full truncated Hilbert space, indicating that the method can recover the physically relevant spectrum without requiring a complete representation of the truncated basis. While the exponential growth of the truncated Hilbert space remains an inherent limitation of Hamiltonian truncation, these results suggest that Krylov-based methods provide a promising strategy for mitigating the computational cost of diagonalization. 

Taken together, the algorithms developed here provide a computational framework for Hamiltonian truncation that improves every stage of the workflow preceding full diagonalization while identifying quantum Krylov methods as a promising direction for future calculations. We expect these developments to facilitate studies at higher truncation scales, more computationally demanding QFTs, and, ultimately, implementations of Hamiltonian truncation on quantum computing hardware.  

\section*{Acknowledgements}

\noindent
The authors would like to thank James Ingoldby for useful discussions. RH and MW are supported by the Institute for Fundamental Theory at the University of Florida. MW is funded by the European Social Fund Plus (ESF Plus) as part of the Margarete von Wrangell Junior Professorship Program. The work of MK, AR and KM is supported in part by the Shelby Endowment for Distinguished Faculty at the University of Alabama. The work of AR and KM is supported in part by Fermilab via Subcontract 731293, in support of DOE Award No.\ DE-SCL0000090 ``HEP AmSC IDA Pilot: Knowledge Extraction'' and DOE Award No.\ DE-SCL0000152 ``USQCD AmSC Infrastructure Provision''. The work of KM is supported in part by the U.S. Department of Energy (DOE) under Award No. DE-SC0026347. 

\appendix

\section{Extrapolation of the \texorpdfstring{$\phi^4$}{phi4} spectrum}
\label{sec:phi4_extrapolation}
Unlike the free theory, the $\phi^4$ theory has no known exact solution.
The extrapolation baselines used in the error decomposition of Sec.~\ref{sec:phi4_results} are therefore obtained by extrapolating the truncated results to $E_\text{max} \to \infty$.
This appendix explains the extrapolation procedure and how the uncertainties are estimated.

\begin{figure}
    \centering
    \includegraphics[width=\linewidth]{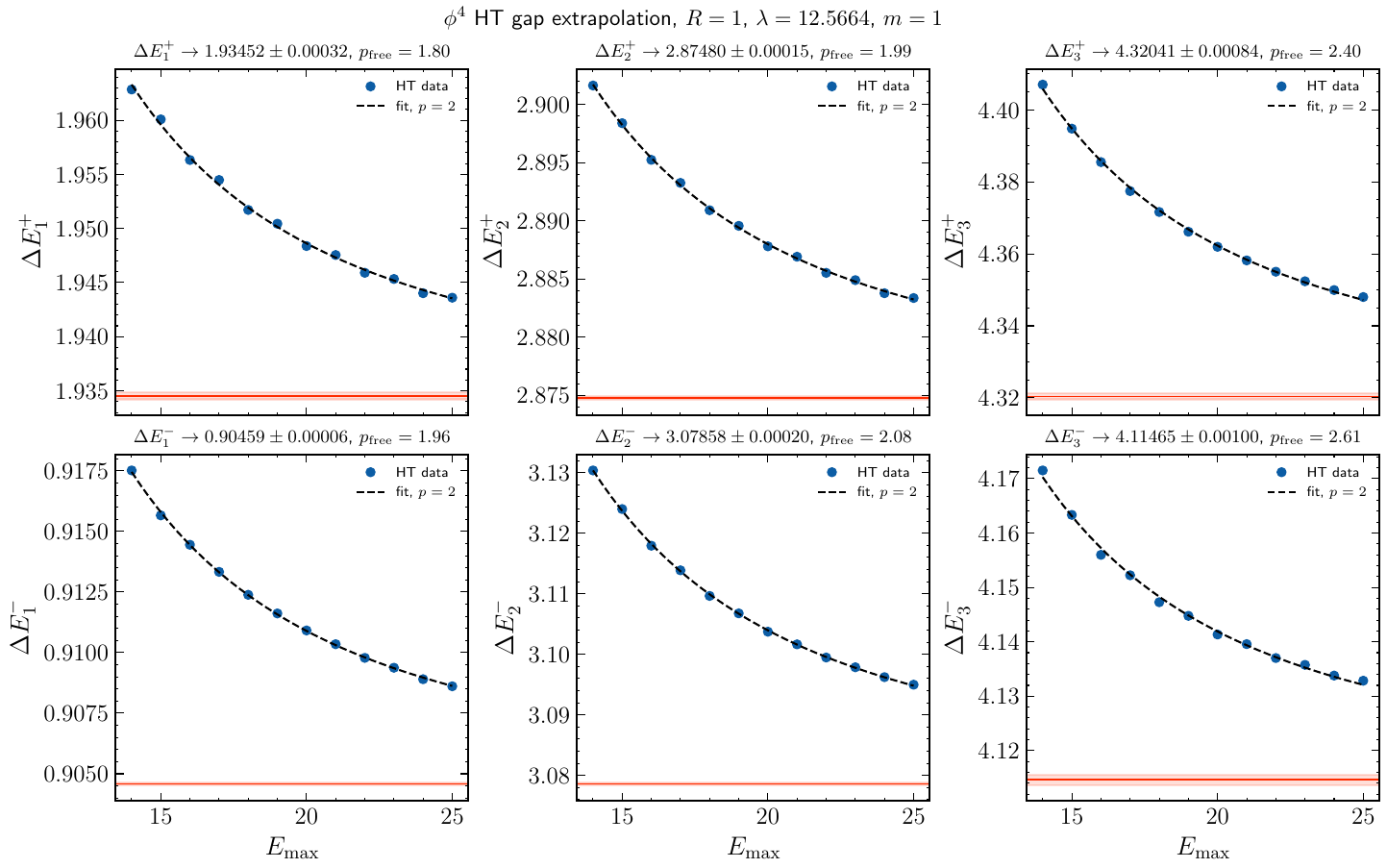}
    \caption{Extrapolation of the $\phi^4$ excitation gaps to $E_\text{max} \to \infty$ at $R = 1$, $\lambda = 4\pi$, $m = 1$ ($m_V=0$).
    Shown are the three lowest gaps in the even (top row) and odd (bottom row) sectors over $E_\text{max} \in [14, 25]$.
    Dashed lines are fits to Eq.~\eqref{eq:gap_extrapolation} with the exponent fixed to the HTET prediction $p = 2$.
    The red line and band mark the extrapolated asymptote $\Delta E_n^{\infty}$ and its total (statistical and systematic) uncertainty.
    Panel titles quote the asymptote and the free-exponent check $p_\text{free}$.
    }
    \label{fig:phi4_extrapolation}
\end{figure}

The main observables are the excitation gaps rather than absolute energies.
The largest truncation error of any single eigenvalue is the state-independent identity-operator contribution~\cite{Demiray:2025zqh}.
This contribution grows with the volume, but it cancels exactly in energy differences.
The gaps therefore converge faster, and they are also the quantities quoted in the literature.
We define $E_0$ as the global (even) ground state and label the gaps by their rank within each sector:
$\Delta E_n^{+} = E_n^{+} - E_0$ is the $n$-th even excitation, and $\Delta E_n^{-} = E_n^{-} - E_0$ is the $n$-th odd level (the odd sector contains no vacuum).
Following Refs.~\cite{Rychkov:2014eea, Cohen:2021erm, Demiray:2025zqh}, each gap is fit to the power law
\begin{equation}
    \Delta E_n(E_\text{max})
    = \Delta E_n^{\infty} + \frac{c_p}{E_\text{max}^{p}} \,,
    \label{eq:gap_extrapolation}
\end{equation}
where the exponent $p$ is \emph{fixed} to the value predicted by the HTET power counting.
This gives $p = 2$ in our case (no LO/NLO corrections or counterterms~\cite{Demiray:2025zqh}), and the fits shown in Fig.~\ref{fig:phi4_extrapolation} use this value.

The absolute ground-state energy is extrapolated separately,
\begin{equation}
    E_0(E_\text{max}) = E_0^{\infty} + \frac{c_0}{E_\text{max}^{p_0}} \,,
    \label{eq:vacuum_extrapolation}
\end{equation}
where the exponent $p_0$ is left free.
The identity-operator error that dominates $E_0$ has its own $E_\text{max}$ scaling, which is different from the scaling of the gaps.
Forcing the gap exponent on the vacuum fit would therefore bias the extrapolation.
Absolute baselines for the comparison with the Krylov Ritz values are then built as $E_n^{\infty} = E_0^{\infty} + \Delta E_n^{\infty}$, where the uncertainties of the two fits are combined quadratically.
Treating the two fits as independent is a conservative choice.

Each extrapolated quantity carries a statistical and a systematic uncertainty, which are added in quadrature and reported in the figure.
The statistical part is the $1\sigma$ uncertainty on the asymptote from the covariance of an unweighted least-squares fit.
Since the truncated eigenvalues carry no independent error bars, this number is only indicative.
The systematic part is estimated by repeating the fit on smaller windows, where up to two of the lowest-$E_\text{max}$ points are dropped.
As a check, each gap is also fit with the exponent left free.
The fitted $p_\text{free}$, reported in the panel titles of Fig.~\ref{fig:phi4_extrapolation}, tests whether the predicted scaling is actually present in the data before the fixed-$p$ extrapolation is trusted.
This protects against levels whose $E_\text{max}$ dependence is dominated by numerical noise instead of the power law, as observed at small volume in Ref.~\cite{Demiray:2025zqh} (Sec.~3.1).

\clearpage
\bibliographystyle{JHEP}
\bibliography{references.bib}

\end{document}